\documentclass[fleqn,usenatbib]{mnras}

\usepackage{newtxtext,newtxmath}

\usepackage[T1]{fontenc}
\usepackage[flushleft]{threeparttable}

\DeclareRobustCommand{\VAN}[3]{#2}
\let\VANthebibliography\thebibliography
\def\thebibliography{\DeclareRobustCommand{\VAN}[3]{##3}\VANthebibliography}

\usepackage{graphicx}	
\usepackage{amsmath}	

\title[Gas-phase Metallicity in the Galactic Center]{{\it Chandra} X-ray Measurement of Gas-phase Heavy Element Abundances in the Central Parsec of the Galaxy}

\author[Z.Q. Hua et al.]{
Ziqian Hua,$^{1,2}$\thanks{E-mail: zqhua@smail.nju.edu.cn}
Zhiyuan Li,$^{1,2}$\thanks{E-mail: lizy@nju.edu.cn}
Mengfei Zhang$^{1,2}$
Zhuo Chen$^{3,4}$
Mark R. Morris$^{3}$
\\
$^{1}$School of Astronomy and Space Science, Nanjing University, Nanjing 210023, China\\
$^{2}$Key Laboratory of Modern Astronomy and Astrophysics (Nanjing University), Ministry of Education, Nanjing 210023, China\\
$^{3}$Department of Physics and Astronomy, University of California, Los Angeles, CA 90095, USA\\
$^{4}$Department of Astronomy, University of Washington, Seattle, WA 98195, USA
}

\date{Accepted XXX. Received YYY; in original form ZZZ}

\pubyear{2015}

\begin{document}
\label{firstpage}
\pagerange{\pageref{firstpage}--\pageref{lastpage}}
\maketitle

\begin{abstract}
Elemental abundances are key to our understanding of star formation and evolution in the Galactic center. Previous work on this topic has been based on infrared (IR) observations, but X-ray observations have the potential of constraining the abundance of heavy elements, mainly through their K-shell emission lines. Using 5.7 Ms {\it Chandra} observations, we provide the first abundance measurement  of Si, S, Ar, Ca and Fe, in four prominent diffuse X-ray features located in the central parsec of the Galaxy, which are the manifestation of shock-heated hot gas. 
A two-temperature, non-equilibrium ionization spectral model is employed to derive the abundances of these five elements. 
In this procedure, a degeneracy is introduced due to uncertainties in the composition of light elements, in particular, H, C and N.
Assuming that the hot gas is H-depleted but C- and N-enriched, as would be expected for a standard scenario in which the hot gas is dominated by Wolf-Rayet star winds, the spectral fit finds a generally subsolar abundance for the heavy elements.
If, instead, the light elements had a solar-like abundance, the heavy elements have a fitted abundance of $\sim$1--2 solar.
The $\alpha$/Fe abundance ratio, on the other hand, is mostly supersolar and insensitive to the exact composition of the light elements.
These results are robust against potential biases due to either a moderate spectral S/N or the presence of non-thermal components.
Implications of the measured abundances for the Galactic center environment are addressed.
\end{abstract}




\section{Introduction} \label{sec:intro}

At a distance of $\sim$8 kpc, the Galactic center offers the best opportunity to unveil the immediate environment of a supermassive black hole (SMBH), commonly known as Sgr A* \citep{2019ApJ...882L..27D, 2020A&A...636L...5G}.
A key component of this environment is the so-called Nuclear Star Cluster (NSC), a dense, spheroidal stellar assembly mainly consisting of $\sim10^7$ low-mass, old stars occupying the inner $\sim10$ parsecs of the Galaxy \citep{2017MNRAS.466.4040F}.
This is also the nearest example of NSCs commonly seen in the local universe {(see recent review by \citealp{2020A&ARv..28....4N}}).
How NSCs form and evolve remains an open question.
Leading scenarios include {\it in situ} star formation, which necessarily survives strong tidal shear by the central SMBH, and migration of star clusters due to the effect of dynamical friction.  
In reality, both channels may work to build up the present-day NSCs.
Regardless, it is generally believed that the formation and evolution of NSCs are closely related to the seeding and growth of SMBHs (e.g., \citealp{2008ApJ...678..116S,2009MNRAS.397.2148G}). 

At the core of the Milky Way NSC, and lying within $\sim$0.5 pc of Sgr A*, is the young nuclear cluster (YNC) containing more than a hundred massive stars, with an estimated age of $\sim$ 4--6 Myr \citep{2010RvMP...82.3121G,2013ApJ...764..155L}.
The existence of massive stars in the vicinity of Sgr A* lends strong support to the {\it in situ} formation channel, since mass segregation is not possible to bring in such young stars either in form of single stars or whole clusters \citep{2001ApJ...546L..39G,2003ApJ...597..312K,2003ApJ...593..352P,2004ApJ...607L.123K}.

Most prominent among the YNC are $\sim$30 Wolf-Rayet (WR) stars, each producing strong winds at a mass loss rate of $\sim10^{-5}\rm~M_\odot~yr^{-1}$ and a terminal velocity of $\sim1000\rm~km~s^{-1}$ \citep{2007A&A...468..233M}.
The mutual collisions of these stellar winds create strong shocks, leading to rapid thermalization of the wind kinetic energy, such as observed in the case of IRS 13E \citep{2020ApJ...897..135Z,2020MNRAS.492.2481W}, a famous infrared (IR) source composed of several closely separated massive stars (including at least two WR stars) belonging to the YNC \citep{2004A&A...423..155M}.
Repeated wind collisions ultimately result in a complex network of hot gas with temperatures $\sim10^{7}$ K pervading the central parsec and beyond, as demonstrated by dedicated hydrodynamic simulations \citep{2004ApJ...604..662R,2006MNRAS.366..358C,2008MNRAS.383..458C, 2017MNRAS.464.4958R,2018MNRAS.478.3544R,2020MNRAS.493..447C,2020ApJ...888L...2C}. 
It is generally thought that this hot gas is responsible for the diffuse thermal X-ray emission  (at photon energies between $\sim 2-8$ keV) detected around Sgr A* \citep{2003ApJ...591..891B,2004ApJ...613..326M,2006MNRAS.367..937W,2013Sci...341..981W}.
The hydrodynamic simulations also predict that Sgr A* is currently fed by this hot gas, at a mass accretion rate only $\sim10^{-8}$ of its Eddington limit, consistent with observational constraints  \citep{2006ApJ...648L.127B,2013Sci...341..981W}. 

The elemental abundances and abundance ratios hold crucial information about the history of star formation and metal enrichment, which is particularly relevant to the origin and evolution of the NSC \citep{2022arXiv221201397C}. 
The $\alpha$-element-to-iron ratio, [$\alpha$/Fe], in particular, is a conventional indicator, as different stellar populations are responsible for the production of $\alpha$-elements and iron on different timescales \citep{1989ARA&A..27..279W,2021A&ARv..29....5M}. 
Measuring the elemental abundances in the Galactic center proves to be challenging, due to severe foreground extinction and source crowding.
Nevertheless, the past decades have witnessed steady progress in the study of elemental abundances in the central parsecs, mostly based on dedicated IR observations.
Pioneering studies focused on the metallicity of the warm, ionized gas, which is concentrated in the {\it mini-spiral} composed of three gas streamers on Keplerian orbits around Sgr A* \citep{2009ApJ...699..186Z}.
\citet{1980ApJ...241..132L} first derived a twice-solar abundance for Ne and Ar, based on mid-IR fine-structure lines from singly and doubly ionized atoms and the hydrogen Br$\gamma$ line.
\citet{1994ApJ...430..236S} further inferred a twice-solar value for Ar but a solar value for Ne, based on emission line strengths averaged over the central half-arcmin and photoionization models, while \citet{2002ApJ...566..880G} found 2$\times$ solar for Ar and Ne at several positions on the mini spiral.
Moreover, \citet{1993ApJ...418..244L} inferred from [Fe III] and [Fe II] lines a $\gtrsim$solar Fe abundance for the {\it mini cavity}, a shell-like feature located at the intersection of the Eastern Arm and Northern Arm of the mini-spiral \citep{1987AIPC..155..127M,1989IAUS..136..443Y,1990Natur.348...45Y}.

More recent studies have focused on the stellar metallicity, primarily targeting  red supergiants and asymptotic giant branch (AGB) stars that are among the brightest stars in the IR.  
Based on absorption lines in the H- and K-band spectra of several red supergiants in the central 2.5 parsec, \citet{2000ApJ...530..307C} and \citet{2000ApJ...537..205R} found a $\sim$solar abundance for Fe. \citet{2007ApJ...669.1011C} and \citet{2009ApJ...694...46D} further derived a supersolar or solar [$\alpha$/Fe] based on absorption lines from O, Mg, Si, Ca and Ti.
\citet{2003ApJ...597..323B} studied more supergiants and AGB stars in the central 5 parsec and preferred [Fe/H]=0. 
Assisted with sensitive Gemini, Keck or VLT observations, \citet{2015ApJ...809..143D,2017MNRAS.464..194F,2020ApJ...894...26T} were able to measure the metallicity of more than 700 fainter stars (mostly red giants), which exhibit a bimodal distribution consisting of both a metal-poor and a metal-rich component, with the metallicity ranging from 1/10 solar to supersolar.
About 6\% of these stars have a low metallicity ([M/H]$< -0.5$).
Several stars with extreme metallicity in \citet{2015ApJ...809..143D} were further analyzed, for which a near-solar [$\alpha$/Fe] was  found \citep{2018ApJ...855L...5D,2022ApJ...925...77B}. 
The existence of metal-poor stars suggests that at least part of the NSC stars have formed {\it ex situ} and migrated to the Galactic center due to dynamical friction.

So far most constraints on the elemental abundances in the central parsecs have been obtained from IR observations, and only a few attempts have been made to constrain the gas-phase metallicity. 
The X-ray spectrum of the diffuse hot gas prevalent in the central parsec primarily consists of a bremsstrahlung continuum\footnote{In principle, free-bound emission and two-photon emission also contribute to the continuum of a thermal plasma. Already included in the spectral models used to fit the observed X-ray spectra (see Section~\ref{subsec:proc}), these two components have a small contribution compared to the bremsstrahlung (i.e., free-free emission). Moreover, both the free-bound and two-photon emissivity have a dependency on the product of ion density and electron density, same as the bremsstrahlung.} and multiple emission lines, typically the K-shell transitions of heavy elements \citep{2004ApJ...613..326M}. In principle, one can make use of this thermal X-ray spectrum to determine or constrain the elemental  abundances, and in turn to shed light on the stellar metallicity of the youngest stellar population in the NSC, provided that the diffuse hot gas within the central parsec is indeed dominated by the winds ejected from the WR stars.    

We take up such a task in this work, utilizing high-resolution and high-sensitivity {\it Chandra} data with a total exposure of 5.7 Ms.
This paper is organized as follows. In Section~\ref{sec:obsinfo}  we describe the {\it Chandra} data and define the targeted diffuse X-ray features in the central parsec. In Section~\ref{sec:model}, we detail the spectral analysis, from which we obtain the first quantitative constraints on the metallicity of the hot gas in the central parsec. In Section~\ref{sec:discussion}, we consider potential caveats in our spectral fit results, in particular, bias due to limited counting statistics or contamination by non-thermal emission. Implications of the measured heavy element abundances are then addressed. 
A summary of this study is provided in  Section~\ref{sec:conclu}.
Throughout this work, we adopt a distance of 8 kpc for the Galactic center. 

\section{Data Preparation and Target Selection} \label{sec:obsinfo}

\subsection{Chandra Data}  \label{subsec:data}
The {\it Chandra} X-ray data utilized in this work are similar to those presented in \citet{2020ApJ...897..135Z}, which result from monitoring observations of Sgr A* and its vicinity over the past two decades.
All observations were taken with the Advanced CCD Imaging Spectrometer (ACIS) and are divided into three groups depending on the CCD in use and the observing mode. 
The ACIS-I group consists of 47 individual observations taken between 1999--2011, with the I3 CCD on-axis; the ACIS-G group consists of 38 observations taken in 2012, with the High Energy Transmission Grating in operation and the S3 CCD on-axis;
the ACIS-S group consists of 41 observations taken between 2014--2019 also with the S3 CCD on-axis. 
Compared to \citet{2020ApJ...897..135Z}, we have added 14 new ACIS-S observations taken since July 2017. We have also omitted 12 ACIS-S observations taken in 2013, during which the outburst of the magnetar, SGR J1745-2900 \citep{2013ApJ...770L..24K}, introduced significant contamination to its surrounding. Similarly, 4 ACIS-S observations taken between February--July 2016 were omitted, due to potential contamination by a bright transient source, {\it Swift} J174540.7-290015 \citep{2016MNRAS.461.2688P}.

\begin{figure*}
	\includegraphics[width= \textwidth]{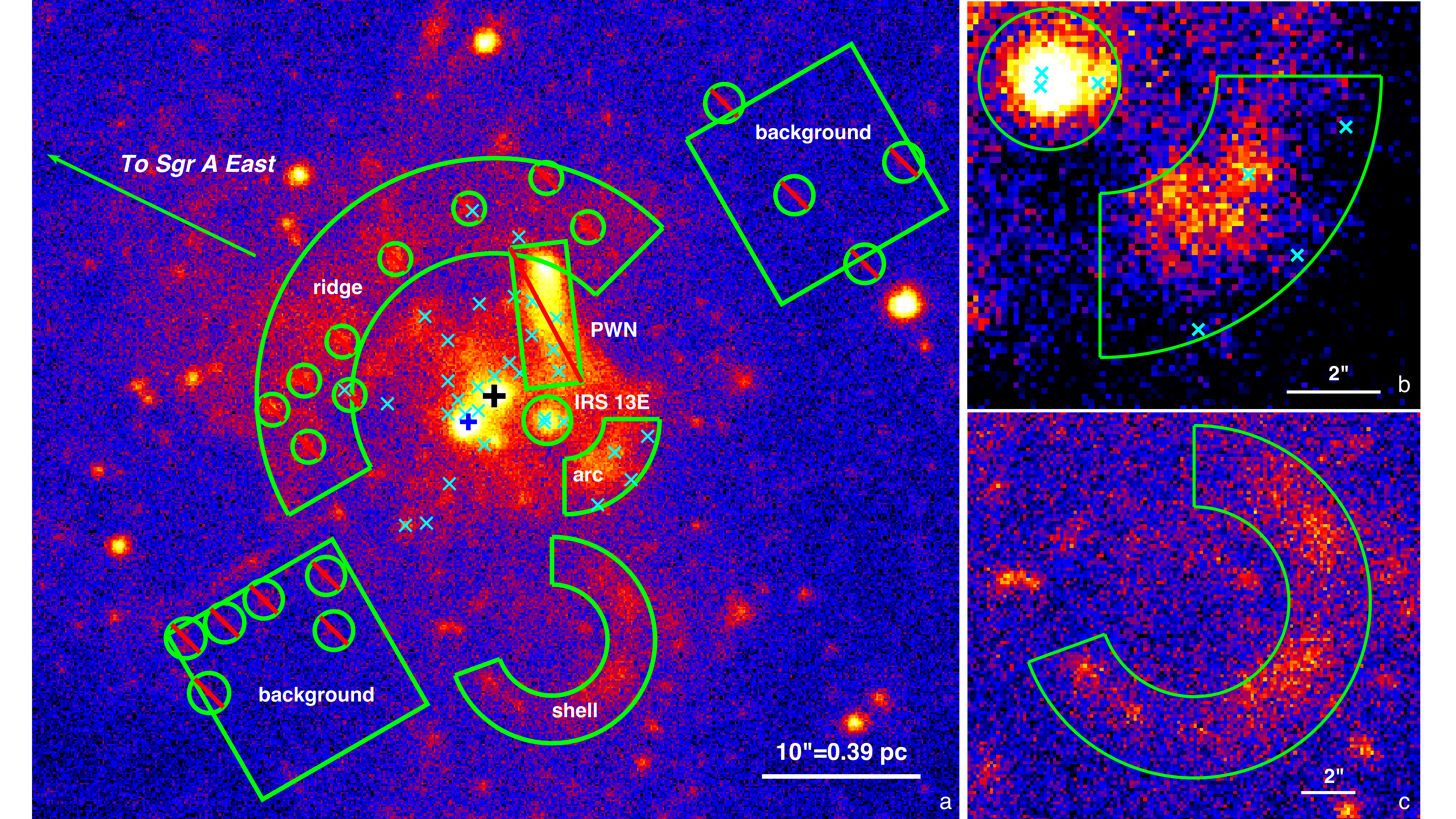}
    \caption{(a) A {\it Chandra}/ACIS 2--8 keV counts image of the Galactic center, combining 122 individual observations. North is up. 
    A binning of 1/4 of the natural ACIS pixel is adopted for better visualization of the diffuse emission. Diffuse features of interest are labeled. Local background regions are outlined by the two green squares. 
    Discrete sources, as well as the candidate PWN, are masked from spectral extraction.
    The position of Sgr A* is marked by a black `+', while the magnetar SGR J1745-2900 is marked by a blue `+'. Known Wolf-Rayet stars, the strong winds of which are thought to produce the diffuse features, are marked by cyan `X's. 
    (b) A zoom-in view of the `arc' and IRS 13E. (c) A zoom-in view of the `shell'.}
    \label{fig:xray_image}
\end{figure*}

A total of 122 observations are thus included. 
The data were downloaded from the public archive and uniformly reprocessed with CIAO 4.12 and CALDB 4.9.2, following the standard procedures detailed in \citet{2020ApJ...897..135Z}. 
The ACIS-I, ACIS-G and ACIS-S groups have a total cleaned exposure of 1.46, 2.96 and 1.27 Ms, respectively. 
Figure~\ref{fig:xray_image} displays the 2--8 keV counts image of the central $60\arcsec \times 50\arcsec$ region combining the 122 observations.
The spectra of particular regions of interest, described in Section~\ref{subsec:region} and indicated in Figure~\ref{fig:xray_image}, were extracted from the individual observations, using the CIAO tool {\it specextract} and co-added for each of the three groups. 
For the ACIS-G observations, we only extracted the non-dispersed spectra from the zeroth-order image.
It is noteworthy that significant differential sensitivity, due to the gradual degradation of the ACIS effective area, occurs primarily for photon energies below $\sim$ 2 keV\footnote{https://cxc.cfa.harvard.edu/cgi-bin/prop\_viewer/build\_viewer.cgi?ea}, thus no significant bias is expected in the coadded spectra of these Galactic center sources, from which only the emission above $\sim$ 2 keV is unextinguished.
Spectra of the local background, outlined in Figure~\ref{fig:xray_image}, were similarly extracted and co-added.
 Known discrete sources in the central parsec \citep{2018ApJS..235...26Z}, as well as the extended source G359.95–0.04, a candidate pulsar wind nebula (PWN; \citealp{2006MNRAS.367..937W}), were masked from the spectral extraction.

\subsection{Diffuse X-ray Features of Interest}
\label{subsec:region}
Four prominent diffuse X-ray features in the central parsec are selected for the following spectral analysis, because (i) they are most likely the manifestations of shock-heated gas, (ii) their X-ray spectra have a sufficiently high signal-to-noise ratio (S/N) for a robust measurement of the metal abundances, and (iii) their relatively high surface brightness minimizes the uncertainty related to the local background.  

The first feature, also the most extended one among the four, is the so-called `X-ray ridge' \citep{2005ApJ...635L.141R},  
an arc-shaped feature spanning a radial range of 0.35--0.6 pc northeast of Sgr A* and an opening angle of $\sim 165^{\circ}$. 
\citet{2005ApJ...635L.141R} proposed that the X-ray ridge originates from an ongoing collision between the collective winds of WR stars around Sgr A*
and the expanding ejecta of Sgr A East, the prominent shell-like radio source that is widely thought to be a supernova remnant (\citealp{1983A&A...122..143E, 2002ApJ...570..671M}). 
This scenario was verified by
\citet{2005ApJ...635L.141R}, and more recently, by Zhang, Li \& Morris (2022, submitted), using numerical simulations. In particular, Zhang et al., using 3-dimensional simulations tailored to trace the hydrodynamic evolution of Sgr A East, showed that the X-ray emission from the ridge should be dominated by the impeded and shock-heated stellar winds, with negligible contribution from the supernova ejecta of Sgr A East.

The second and also the most compact feature is known as IRS 13E, originally identified as a bright IR source \citep{2004A&A...423..155M} and found to exhibit thermal X-ray emission by early {\it Chandra} observations \citep{2006MNRAS.367..937W}. 
Located at just 3\farcs5 southwest of Sgr A*, IRS 13E in fact consists of several closely spaced massive stars, including two or three WR stars (\citealp{2006ApJ...643.1011P,2007A&A...468..233M}; Figure~\ref{fig:xray_image}b).
\citet{2000A&A...361L..13C} proposed that the thermal X-ray emission is induced by the colliding winds from the massive stars belonging to IRS 13E.
\citet{2020ApJ...897..135Z} verified this scenario (see also \citealp{2020MNRAS.492.2481W}), finding good agreement between the {\it Chandra} spectrum and a synthetic spectrum based on hydrodynamic simulations tailored to match the physical conditions of the WR stars in IRS 13E.
Following \citet{2020ApJ...897..135Z}, we use a 1\farcs5-radius circle to enclose the slightly extended X-ray emission from IRS 13E (Figure~\ref{fig:xray_image}b).

A third feature, hereafter dubbed the `arc', is an arc-shaped feature located at $\sim3\arcsec$ southwest of IRS 13E (Figure~\ref{fig:xray_image}b). This feature was already noticeable in the early {\it Chandra} observations (e.g., \citealp{2003ApJ...591..891B,2006MNRAS.367..937W}), but to our knowledge, its X-ray properties have not been quantitatively studied. 
The arc curves toward the north and gradually joins the long tail of G359.95-0.04.
We extract the spectrum of the arc from its southern portion  (outlined by a wedge with a thickness of 3\arcsec\ and an opening angle of $90^{\circ}$), which avoids potential contamination of the non-thermal X-ray emission from G359.95-0.04. 
While the exact nature of the arc is unclear thus far, the presence of at least four WR stars in this region (Figure~\ref{fig:xray_image}b), including the Allen-Forrest star AF NW which was proposed to be correlated with this X-ray feature \citep{2003ApJ...591..891B}, strongly suggests that it is also due to shock-heated stellar winds. 
Indeed, a feature highly resembling the arc is clearly seen in the synthesized X-ray image from the hydrodynamic simulation of \citet[][figure 8 therein]{2018MNRAS.478.3544R}, which was based on realistic positions and orbits of the WR stars.

The last feature, hereafter called the `shell', is located south of the arc and has a morphology reminiscent of a partial shell, with its northeastern quadrant showing little excess to the local background (Figure~\ref{fig:xray_image}c).
There are no known massive stars (WR or O-type) on or near the shell, making its relation to stellar winds less certain. Nevertheless, we include this feature in the following spectral analysis to shed light on its origin. 
We use a wedge with inner-to-outer radii of 3\farcs5--6\farcs5 and an opening angle of $250^{\circ}$ to extract its spectrum.

For the following spectral analysis, we adopt two background regions, as shown in Figure~\ref{fig:xray_image}, which are sufficiently close to the four diffuse features yet sufficiently large to represent the local diffuse background.
The background, which is mainly composed of unresolved stellar objects, in particular cataclysmic variables \citep{2018ApJS..235...26Z},  accounts for 10--40\% of the total flux in the source regions.
Our test indicates that the exact choice of the background region has little effect on the spectral fit results, thanks to the relatively high surface brightness of the four features.

\section{X-ray Spectral Analysis} \label{sec:model}
\subsection{Procedures}
\label{subsec:proc}
\begin{figure*}
	\centering
	\includegraphics[width=0.48\textwidth]{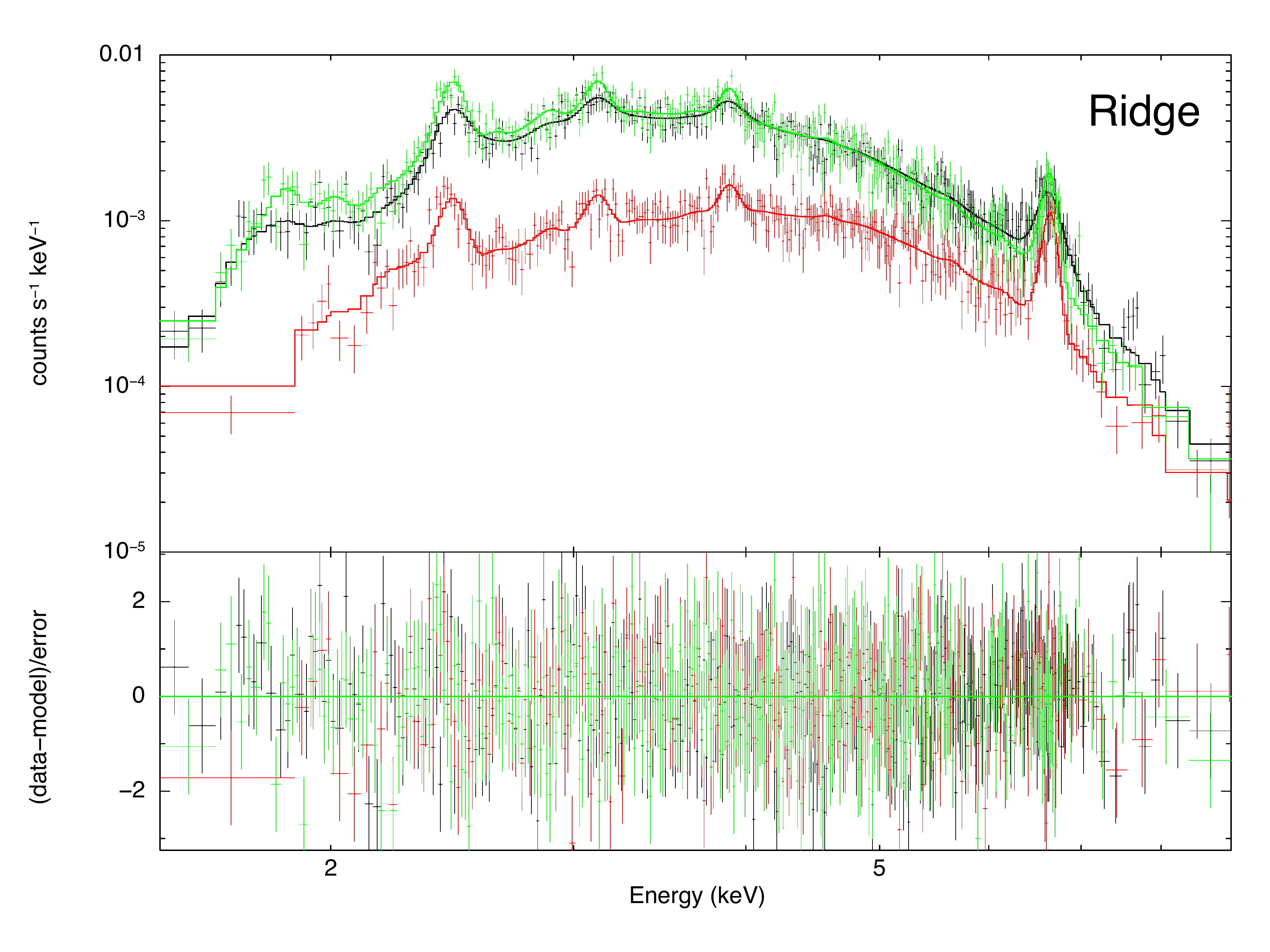}
	\includegraphics[width=0.48\textwidth]{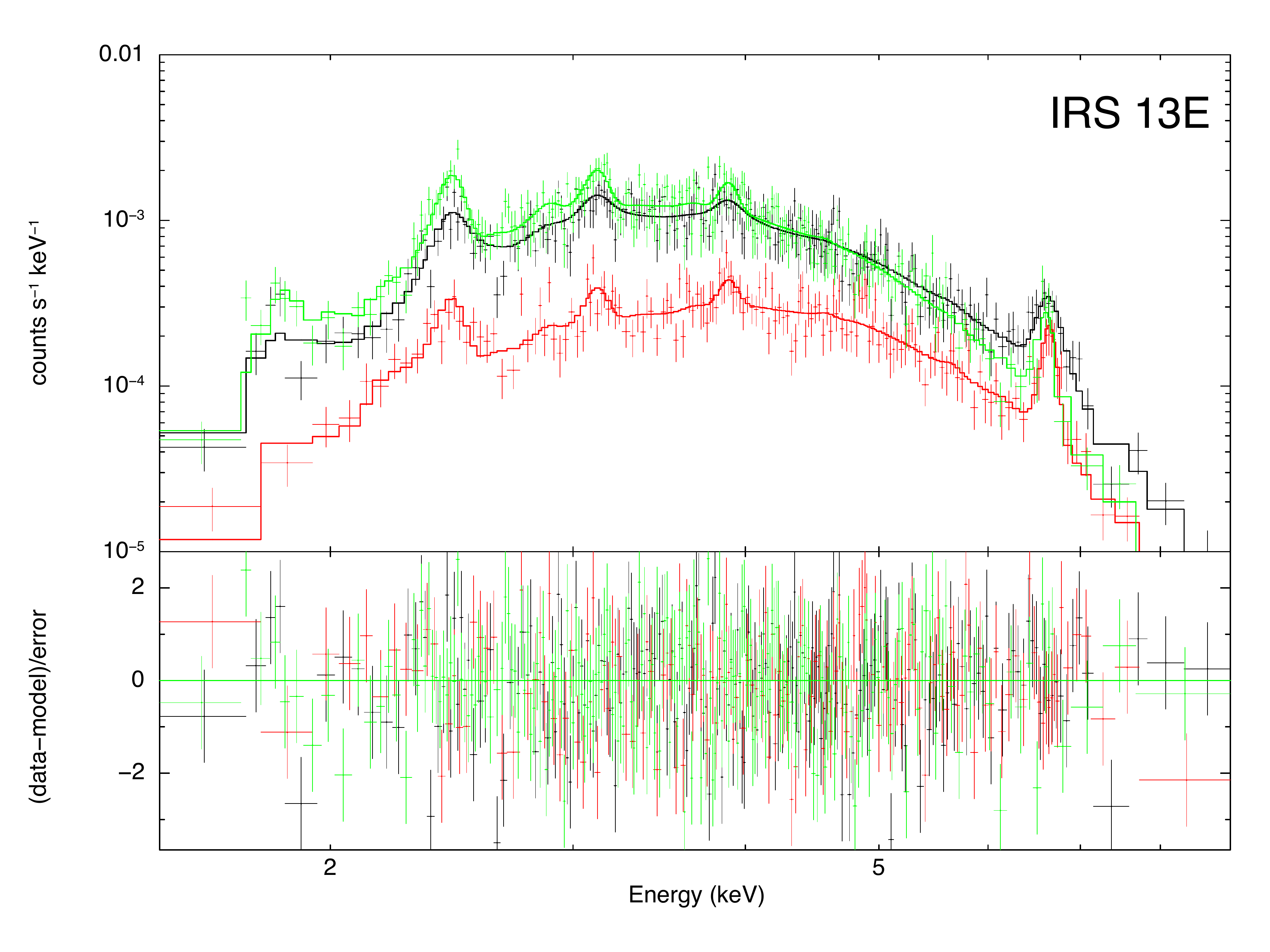}
	\includegraphics[width=0.48\textwidth]{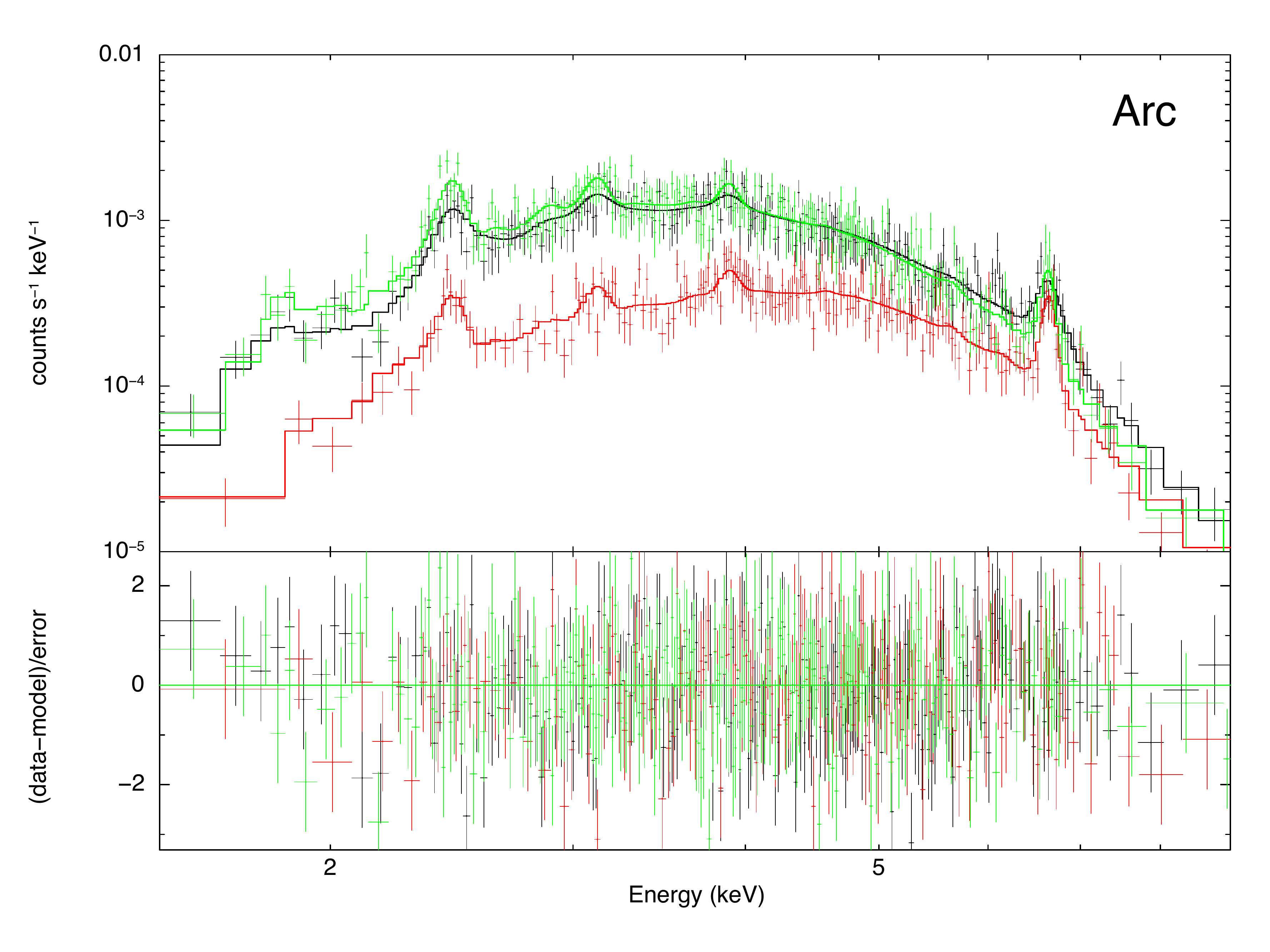}
	\includegraphics[width=0.48\textwidth]{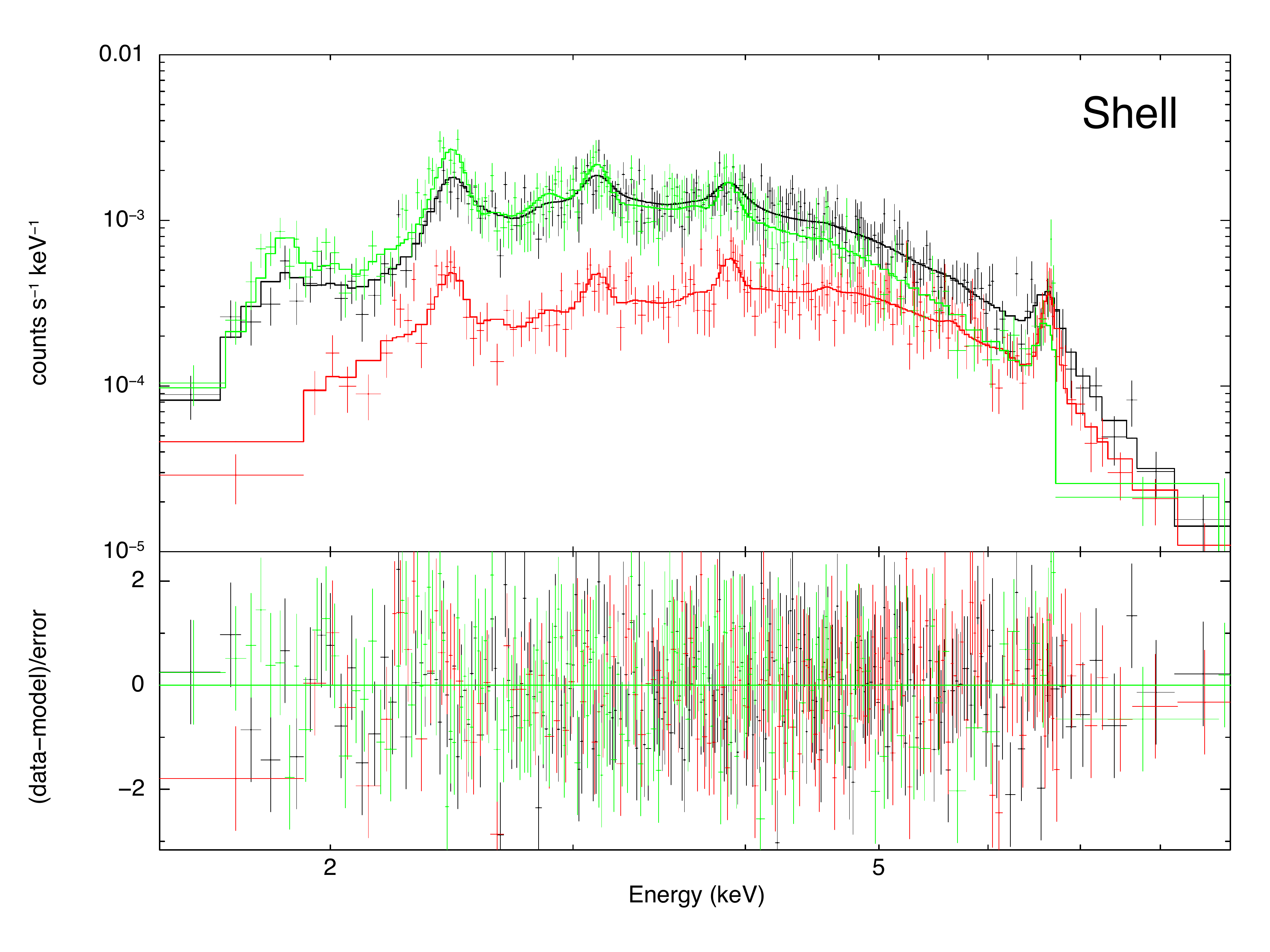}	
	\caption{Spectra of the four diffuse X-ray features. The ACIS-I, -G and -S spectra are shown in black, red and green, respectively, all adaptively binned to have at least 10 counts and a S/N greater than 3 per bin. The  best-fit hydrogen-depleted model, {\it TBABS*DUSTSCAT*(VNEI+VNEI)} in Xspec, with $Z_{\rm H}$=0.1, $Z_{\rm C}$=4.0 and $Z_{\rm N}$=3.5, is shown by the solid curves. The bottom panels show the relative residuals. The error bars are at the 1$\sigma$ level. See text for details.}
	\label{fig:spectra}
\end{figure*}

The background-subtracted spectra of the four diffuse X-ray features are shown in Figure~\ref{fig:spectra}.
Various emission lines are clearly present in all four spectra, most prominently from the $\alpha$ elements Si, S, Ar and Ca, as well as from Fe.
Spectral fitting is performed with Xspec v12.11 \citep{1996ASPC..101...17A}, employing the $C$-statistic, which is more immune to biased parameter estimation than the $\chi^2$-statistic, especially in the regime of moderate counts \citep{2009ApJ...693..822H}, and does not require substantial spectral binning so that spectral features can be best preserved. 
An energy range of 1.5-9.0 keV is considered. Photons with energy $\lesssim$1.5 keV are expected to be fully obscured by the Galactic foreground, while the particle background dominates above 9 keV.
For illustration purposes, we present in the figures the adaptively binned spectra, which have an S/N greater than 3 and a minimum of 10 counts per bin.

As mentioned in Section~\ref{subsec:region}, the diffuse features are most likely the manifestation of shock-heated WR star winds, which should produce an X-ray spectrum characteristic of an optically thin, collisional plasma.
It is also expected that the ionization state of the plasma deviates from collisional ionization equilibrium (CIE), as the result of recent shock perturbation. This has been demonstrated in the case of IRS 13E \citep{2020ApJ...897..135Z}, who found that a single-temperature (1-T), non-equilibrium ionization (NEI) plasma model provides a significantly better fit than a 1-T CIE model. 
Therefore, we start with a 1-T NEI model with variable abundances ({\it vnei} in Xspec). 
However, this model, while providing an apparently acceptable value of $C$/d.o.f., leaves significant deviations at certain energies, especially near the low-energy and high-energy ends.
This is a sign that the underlying plasma has an intrinsic temperature distribution, which is again expected for a post-shock gas. 
Hence, we consider two alternative models to capture this behavior. 

The first model, through a custom implementation in Xspec, is a CIE plasma (based on the APEC model of \citealp{2001ApJ...556L..91S}) with a log-normal temperature distribution (e.g., \citealp{2015ApJ...812..130G}), which in principle can better describe the broadband continuum due to the extra degree of freedom in the temperature distribution. 
However, it is found that this log-normal temperature model cannot simultaneously account for the broadband continuum and the relative strength of the forbidden transition of the helium-like Fe (i.e., Fe XXV) K$\alpha$ triplet around 6.7 keV.
The latter is likely an effect of NEI, which, however, is ill-defined with the log-normal temperature model. 

Therefore, we settle on the second model, which is a two-temperature (2-T) NEI plasma ({\it vnei+vnei} in Xspec, \citealp{2001ApJ...548..820B}). The abundances and the ionization timescale ($\tau$) are assumed to be the same between the two NEI components, but the normalizations are allowed to be different.  
This model can simultaneously account for the broadband continuum and any NEI effects with the emission lines. 
It turns out that the 2-T NEI model provides an improved fit compared to the 1-T NEI model or the log-normal temperature model. 
Below we shall focus on the discussion of this model.


The 2-T NEI model is subject to Galactic foreground absorption, for which the Tuebingen-Boulder interstellar medium (ISM) absorption ({\it TBABS} in Xspec) is employed, with the elemental abundance standard of \citet{2000ApJ...542..914W}.\footnote{The abundance standard of \citet{2000ApJ...542..914W}, which characterizes the ISM, is the default abundance standard to employ the {\it TBABS} model. To convert to the more conventional solar abundance standard of \citet{2009ARA&A..47..481A}, a multiplying factor of 1.15/0.89/1.12/0.57/0.93/1.02/0.72/0.85 should be respectively applied for He/C/N/Si/S/Ar/Ca/Fe.} 
Due to the expected large absorption column density toward the Galactic center ($N_{\rm H} \sim 10^{23}\rm~cm^{-2}$), foreground dust scattering is relevant.
Following \citet{2013ApJ...779..154L}, we include the spectral hardening effect of dust scattering in the spectral fit, i.e., a multiplying factor of ${\rm exp}(-0.486 E^{-2}N_{\rm H})$, where $E$ is the photon energy in units of keV and $N_{\rm H}$ is in units of $10^{22}\rm~cm^{-2}$ \citep{1995A&A...293..889P}. 
We note that the inclusion of dust scattering mainly affects the derived values of $N_{\rm H}$ and the plasma temperature, but has little effect on the metal abundances. 

We focus on the abundances of Si, S, Ar, Ca and Fe, which are set as free parameters during the spectral fit. 
We note that Si presents itself mainly by the Si XIII K$\alpha$ line at $\sim$1.8 keV, despite the strong absorption. 
The lighter metals, including C, N, O, Ne, Mg and Al, have their prominent lines well below 1.5 keV, thus their abundances cannot be well constrained with the present spectra. 
In the {\it vnei} model (or any other model of a hot plasma), 
light elements, in particular H, He, C and N, dominate the continuum via their bremsstrahlung emission, which scales with $n_{\rm e}\Sigma(n_{\rm i}A_{\rm i}^2) \propto n_{\rm e}n_{\rm H}\Sigma(f_{\rm i}Z_{\rm i}A_{\rm i}^2)$, where $n_{\rm e}$ and $n_{\rm i}$ are the electron and ion density, $A_{\rm i}$ is the atomic number of the ion, $f_{\rm i}$ denotes the ratio of the number density of element $i$ to hydrogen in the abundance standard of \citet{2000ApJ...542..914W} and $Z_{\rm i}$ denotes the abundance of element $i$.
Thus an ambiguity in the amount of these light elements would necessarily result in a correlated uncertainty in the abundance of the heavy elements, which is measured via their line emission that scales with $n_{\rm e}n_{\rm j} \propto n_{\rm e}n_{\rm H}f_{\rm j}Z_{\rm j}$.

One would expect that the elemental abundances of the diffuse X-ray features largely reflect the chemical composition of the WR stars, which is highly hydrogen-depleted but C- and N-enriched \citep{2007ARA&A..45..177C}. 
Indeed, \citet{2007A&A...468..233M} were able to measure the number density ratio of C to He as well as the N mass fraction in about half of the WR stars detected around Sgr A*, based on their IR spectra.
Averaging over these WR stars, we estimate a mean abundance of $Z_{\rm C}$ = 4.4 and $Z_{\rm N}$ = 3.8. 
Taking advantage of this knowledge, we consider the following two fiducial compositions of H, He, C and N in our spectral modeling: 

(i) The hydrogen-depleted case: $Z_{\rm H}$ = 0.1, $Z_{\rm He}$ = 1, $Z_{\rm C}$ = 4.0, $Z_{\rm N}$ = 3.5\footnote{Under the standard of \citet{2000ApJ...542..914W}, the number density ratio of He, C, N to H is $9.77\times10^{-2}$, $2.40\times10^{-4}$ and $7.59\times10^{-5}$, respectively.}.
This implicitly assumes that the C- and N- enriched winds from the individual WR stars are well mixed in the four diffuse features. 
Here $Z_{\rm H}$ = 0.1 {\it should be understood as a hydrogen number density 10\% of the reference standard of \citet{2000ApJ...542..914W}}.
A 10\% hydrogen accounts for any residual hydrogen in the WR star winds as well as potential contribution from other stars in the vicinity of Sgr A* (see more discussions in Section~\ref{subsec:Hsupply}).
It now makes sense to take He as the reference element.
In this case the contribution of H, C, N to the bremsstrahlung is 25.6\%, 8.8\% and 3.3\%, respectively, with respect to that of He. As mentioned in Section~\ref{sec:intro}, the free-bound emission should follow a similar proportion among the different elements, which is taken into account in the spectral fit.

(ii) The hydrogen-normal case: $Z_{\rm H}$ = $Z_{\rm He}$ = $Z_{\rm C}$ = $Z_{\rm N}$ = 1. 
In this case the contribution of H, C, N to the bremsstrahlung is 255.9\%, 2.2\% and 1.0\%, respectively, with respect to that of He.

While the first case is considered physically more plausible, the second case is more conventional in X-ray spectral analyses of the hot ISM and allows us to assess the uncertainty related to the possible range of abundances in H, C and N, hence the systematic uncertainty in the continuum level.    
In both cases, the abundance of other elements, including O, Ne, Mg, Al and Ni, is fixed at solar. Since these elements make only a minor contribution to the continuum, any uncertainty related to their abundances should be readily absorbed into the results covered by the two fiducial cases.  


The ACIS-I, ACIS-G and ACIS-S spectra of a given feature, which have a comparable degree-of-freedom, are jointly fitted, with the column density, ionization timescale, and metal abundances linked between the three spectra.
The ionization timescale and metal abundances are further linked between the low- and high-temperature components.
The temperature and normalization of the low-temperature component are also linked between the three spectra. Allowing these latter two parameters to vary does not lead to significant improvement of the fit.
We allow the temperature and normalization of the high-temperature component to vary between the three spectra, which accounts for a mild difference in the overall spectral shape and leads to a significantly improved fit according to the $F$-test. 

In short, the 2-T {\it vnei} model has fifteen free parameters, including the absorption column density, ionization timescale, the abundances of Si, S, Ar, Ca and Fe, the temperature and normalization of the cooler plasma component, and three temperatures and three normalizations of the hotter plasma.
It turns out that this model provides a reasonable fit to all four spectra (Figure~\ref{fig:spectra}), and the hydrogen-depleted and hydrogen-normal models provide a quite comparable $C/d.o.f.$ for all four features.
The fit results are summarized in Table~\ref{tab:table1}.

For both the hydrogen-depleted and hydrogen-normal cases, we also present in Table~\ref{tab:table1} the unabsorbed 2--8 keV luminosity, the mean hydrogen density and total gas mass for each feature, based on the best-fit parameters of the ACIS-I spectra. 
To derive the mean hydrogen density and total gas mass, we have adopted the usual assumptions that
the low-temperature and high-temperature components are in pressure equilibrium and that they sum up to a volume filling factor of unity. The volume of each feature is estimated as $V=S^{1.5}$, where $S$ is the projected area of the spectral extraction region. 90\% uncertainties of these parameters are estimated using the Xspec function {\it fakeit} based on a bootstrapping method.

\begin{table*}
	\centering
	\renewcommand\arraystretch{1.3}
	\setlength\tabcolsep{10pt}
	\caption{X-ray Spectral Fit Results\label{tab:table1}}
	\begin{threeparttable}

	\begin{tabular}{llccccc} 
	\hline
	\hline
	Model & Parameter & ridge & IRS 13E & arc & shell\\
	\hline

 $Z_{\rm H}$=0.1, $Z_{\rm C}$=4.0, $Z_{\rm N}$=3.5 &  $\rm{N_H}$ ($10^{22}\ \rm{cm}^{-2}$)  & $11.47_{-0.52}^{+0.60}$ & $14.62_{-0.70}^{+1.85}$ & $13.14_{-1.37}^{+1.30}$ & $12.13_{-1.13}^{+1.41}$ \\
 & $\rm{Si}$ & $0.35_{-0.13}^{+0.15}$ & $1.02_{-0.50}^{+1.05}$ & $0.59_{-0.31}^{+0.44}$ & $0.69_{-0.31}^{+0.38}$ \\
 & $\rm{S}$ & $0.45_{-0.05}^{+0.05}$ & $0.58_{-0.10}^{+0.13}$ & $0.50_{-0.09}^{+0.10}$ & $0.66_{-0.10}^{+0.15}$ \\
 & $\rm{Ar}$ & $0.57_{-0.10}^{+0.11}$ & $0.70_{-0.17}^{+0.19}$ & $0.52_{-0.17}^{+0.19}$ & $0.84_{-0.23}^{+0.34}$ \\
 & $\rm{Ca}$ & $0.94_{-0.17}^{+0.20}$ & $0.98_{-0.27}^{+0.31}$ & $0.70_{-0.22}^{+0.24}$ & $1.01_{-0.27}^{+0.28}$ \\
 & $\rm{Fe}$ & $0.55_{-0.08}^{+0.08}$ & $0.54_{-0.13}^{+0.16}$ & $0.37_{-0.07}^{+0.09}$ & $0.34_{-0.09}^{+0.08}$ \\
 & $\tau$ ($10^{11}\ \rm{cm}^{-3}\ \rm{s}$)  & $0.87_{-0.12}^{+0.16}$ & $1.21_{-0.33}^{+0.65}$ & $0.89_{-0.16}^{+0.24}$ & $0.99_{-0.25}^{+0.42}$ \\
 & $\rm{kT_{l}}$ (keV) & $1.16_{-0.19}^{+0.22}$ & $0.81_{-0.25}^{+0.25}$ & $0.86_{-0.19}^{+0.23}$ & $0.55_{-0.16}^{+0.17}$ \\
 & $\rm{Norm_{l}}$ ($10^{-3}\ \rm{cm}^{-5}$) & $7.24_{-1.99}^{+3.08}$ & $5.50_{-2.71}^{+9.01}$ & $3.60_{-2.08}^{+4.16}$ & $9.35_{-6.94}^{+22.60}$ \\
 & $\rm{kT_{h,I}}$ (keV) & $4.88_{-1.09}^{+1.02}$ & $3.00_{-0.74}^{+1.50}$ & $4.13_{-0.89}^{+1.11}$ & $2.83_{-0.47}^{+0.58}$ \\
 & $\rm{Norm_{h,I}}$ ($10^{-3}\ \rm{cm}^{-5}$) & $1.06_{-0.31}^{+0.49}$ & $0.53_{-0.27}^{+0.43}$ & $0.56_{-0.15}^{+0.28}$ & $0.94_{-0.14}^{+0.33}$ \\
 & $\rm{kT_{h,G}}$ (keV) & $10.79_{-5.19}^{+25.10}$ & $3.41_{-1.02}^{+2.72}$ & $5.44_{-1.52}^{+2.66}$ & $4.29_{-0.96}^{+1.78}$ \\
 & $\rm{Norm_{h,G}}$ ($10^{-3}\ \rm{cm}^{-5}$) & $0.55_{-0.16}^{+0.36}$ & $0.37_{-0.21}^{+0.37}$ & $0.44_{-0.11}^{+0.20}$ & $0.61_{-0.18}^{+0.23}$ \\
 & $\rm{kT_{h,S}}$ (keV) & $5.47_{-1.14}^{+2.39}$ & $2.08_{-0.35}^{+0.48}$ & $4.03_{-0.88}^{+1.03}$ & $2.22_{-0.35}^{+0.43}$ \\
 & $\rm{Norm_{h,S}}$ ($10^{-3}\ \rm{cm}^{-5}$) & $0.98_{-0.29}^{+0.46}$ & $0.86_{-0.39}^{+0.61}$ & $0.56_{-0.15}^{+0.29}$ & $0.98_{-0.31}^{+0.44}$ \\

 & $C/d.o.f.$ & 1663.77/1517 & 1328.69/1252 & 1365.11/1399 & 1419.93/1470 \\
  & $L_{2-8} (10^{33}\ \rm{erg}\ \rm{s}^{-1})$  & $8.98^{+0.09}_{-0.09}$ & $4.03^{+0.06}_{-0.06}$ & $3.32^{+0.05}_{-0.05}$ & $3.56^{+0.06}_{-0.05}$ \\
  & $\bar{n}_{\rm H}$ ($\rm cm^{-3}$) & $5.8_{-0.8}^{+1.4}$ & $58.8_{-16.0}^{+42.1}$ & $11.8_{-3.0}^{+6.6}$ & $12.9_{-4.5}^{+14.2}$ \\
  & $M_{\rm gas}$ ($10^{-2}\rm~M_{\odot}$) & $9.8_{-1.4}^{+2.4}$ & $0.8_{-0.2}^{+0.5}$ & $2.2_{-0.6}^{+1.2}$ & $4.9_{-1.7}^{+5.4}$ \\
	\hline
	\hline

$Z_{\rm H}$=1, $Z_{\rm C}$=1, $Z_{\rm N}$=1   & $\rm{N_H}$ ($10^{22}\ \rm{cm}^{-2}$) & $11.21_{-0.64}^{+0.57}$ & $14.37_{-1.31}^{+1.76}$ & $12.74_{-1.21}^{+1.33}$ & $12.17_{-1.89}^{+1.42}$ \\
 & $\rm{Si}$ & $0.57_{-0.22}^{+0.28}$ & $1.54_{-0.70}^{+1.01}$ & $0.84_{-0.44}^{+0.34}$ & $0.94_{-0.29}^{+0.51}$ \\
 & $\rm{S}$ & $0.79_{-0.09}^{+0.12}$ & $0.92_{-0.16}^{+0.19}$ & $0.84_{-0.15}^{+0.18}$ & $1.04_{-0.20}^{+0.18}$ \\
 & $\rm{Ar}$ & $1.05_{-0.10}^{+0.20}$ & $1.17_{-0.28}^{+0.31}$ & $0.99_{-0.33}^{+0.38}$ & $1.79_{-0.54}^{+0.56}$ \\
 & $\rm{Ca}$ & $1.82_{-0.37}^{+0.37}$ & $1.72_{-0.52}^{+0.60}$ & $1.41_{-0.44}^{+0.24}$ & $2.06_{-0.56}^{+0.62}$ \\
 & $\rm{Fe}$ & $1.16_{-0.18}^{+0.17}$ & $1.05_{-0.26}^{+0.33}$ & $0.77_{-0.15}^{+0.18}$ & $0.71_{-0.19}^{+0.23}$ \\
 & $\tau$ ($10^{11}\ \rm{cm}^{-3}\ \rm{s}$) & $0.89_{-0.12}^{+0.23}$ & $1.22_{-0.32}^{+0.70}$ & $0.91_{-0.17}^{+0.26}$ & $0.98_{-0.27}^{+0.45}$ \\
 & $\rm{kT_{l}}$ (keV) & $1.19_{-0.18}^{+0.26}$ & $0.85_{-0.24}^{+0.23}$ & $0.86_{-0.23}^{+0.24}$ & $0.47_{-0.10}^{+0.62}$ \\
 & $\rm{Norm_{l}}$ ($10^{-3}\ \rm{cm}^{-5}$) & $3.79_{-1.21}^{+1.45}$ & $3.07_{-1.51}^{+6.84}$ & $2.05_{-1.14}^{+3.27}$ & $14.70_{-13.70}^{+62.00}$ \\
 & $\rm{kT_{h,I}}$ (keV) & $4.84_{-1.04}^{+2.05}$ & $3.08_{-0.80}^{+1.03}$ & $4.08_{-0.86}^{+0.97}$ & $2.83_{-0.46}^{+0.94}$ \\
 & $\rm{Norm_{h,I}}$ ($10^{-3}\ \rm{cm}^{-5}$) & $0.49_{-0.17}^{+0.23}$ & $0.25_{-0.14}^{+0.24}$ & $0.27_{-0.07}^{+0.12}$ & $0.47_{-0.17}^{+0.18}$ \\
 & $\rm{kT_{h,G}}$ (keV) & $10.89_{-5.33}^{+34.29}$ & $3.56_{-1.24}^{+3.52}$ & $5.43_{-1.53}^{+1.78}$ & $4.33_{-0.93}^{+2.45}$ \\
 & $\rm{Norm_{h,G}}$ ($10^{-3}\ \rm{cm}^{-5}$) & $0.24_{-0.08}^{+0.17}$ & $0.17_{-0.11}^{+0.20}$ & $0.20_{-0.05}^{+0.10}$ & $0.29_{-0.12}^{+0.11}$ \\
 & $\rm{kT_{h,S}}$ (keV) & $5.48_{-1.31}^{+3.14}$ & $2.12_{-0.37}^{+0.43}$ & $4.01_{-0.86}^{+0.84}$ & $2.26_{-0.34}^{+0.34}$ \\
 & $\rm{Norm_{h,S}}$ ($10^{-3}\ \rm{cm}^{-5}$) & $0.45_{-0.15}^{+0.21}$ & $0.43_{-0.20}^{+0.36}$ & $0.27_{-0.06}^{+0.13}$ & $0.49_{-0.21}^{+0.22}$ \\
  & $C/d.o.f.$ & 1667.83/1517 & 1327.57/1252 & 1364.29/1399 & 1417.57/1470 \\
& $L_{2-8} (10^{33}\ \rm{erg}\ \rm{s}^{-1})$  & $8.61^{+0.08}_{-0.09}$ & $3.81^{+0.06}_{-0.06}$ & $3.08^{+0.05}_{-0.04}$ & $3.58^{+0.06}_{-0.06}$ \\
  & $\bar{n}_{\rm H}$ ($\rm cm^{-3}$) & $21.4_{-4.6}^{+4.2}$ & $221.9_{-46.9}^{+210.6}$ & $44.7_{-11.5}^{+33.4}$ & $82.5_{-32.6}^{+155.5}$ \\
  & $M_{\rm gas}$ ($10^{-2}\rm~M_{\odot}$) & $10.1_{-2.1}^{+2.2}$ & $0.8_{-0.1}^{+0.8}$ & $2.3_{-0.6}^{+1.7}$ & $8.8_{-3.8}^{+16.4}$ \\
    \hline
    \hline
	\end{tabular}
	\begin{tablenotes}
	\small
	\item
	Notes:
	The upper and low panels show the best-fit parameters with the hydrogen-depleted and hydrogen-normal cases, both using an absorbed two-temperature {\it vnei} model. The subscripts `l' and `h' denote the low and high temperature components. The abundances and ionization timescale ($\tau$) are linked between the low- and high-temperature components. The temperature and normalization of the high-temperature component are allowed to vary between the three spectra,  which are denoted by subscripts `I', `G' and `S'. Reported errors are at the 90\% confidence level. $L_{2-8}$ denotes the unabsorbed 2--8 keV X-ray luminosity, while $\bar{n}_{\rm H}$ and $M_{\rm gas}$ denote the mean  hydrogen density and total gas mass, based on the best-fit parameters of the ACIS-I spectra. See text for details on the spectral models.
	\end{tablenotes} 
\end{threeparttable}
\end{table*}

\subsection{Results}
The best-fit foreground column density is found to be $N_{\rm H} \sim (11-15)\times10^{22}\rm~cm^{-2}$, 
with $\lesssim20\%$ variation 
among the four diffuse features. 
These values are expected for their locations in the central parsec, and in the case of IRS13E is in good agreement with previous work \citep{2020ApJ...897..135Z}.
The best-fit temperature is found to be 0.5--1 keV for the low-temperature component and 3--5 keV for the high-temperature component.
For a given feature, the best-fit temperatures do not change significantly between the two cases of chemical composition, neither does the column density. 
This is understood because the temperature is mainly controlled by the overall shape of the bremsstrahlung continuum, but is insensitive to the metal abundances.



\begin{figure*}
	\centering
	\includegraphics[width=0.48\textwidth]{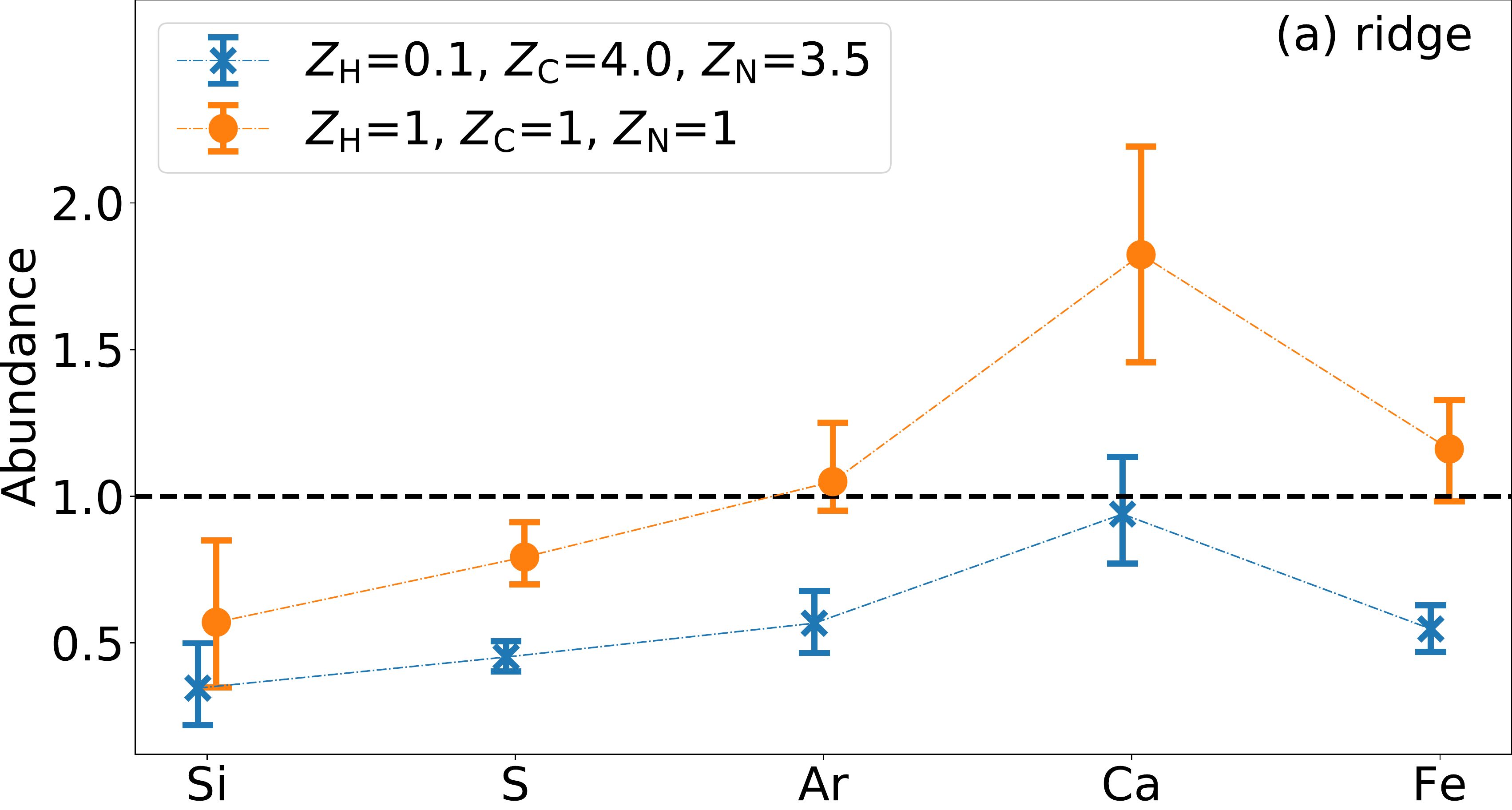}
	\includegraphics[width=0.48\textwidth]{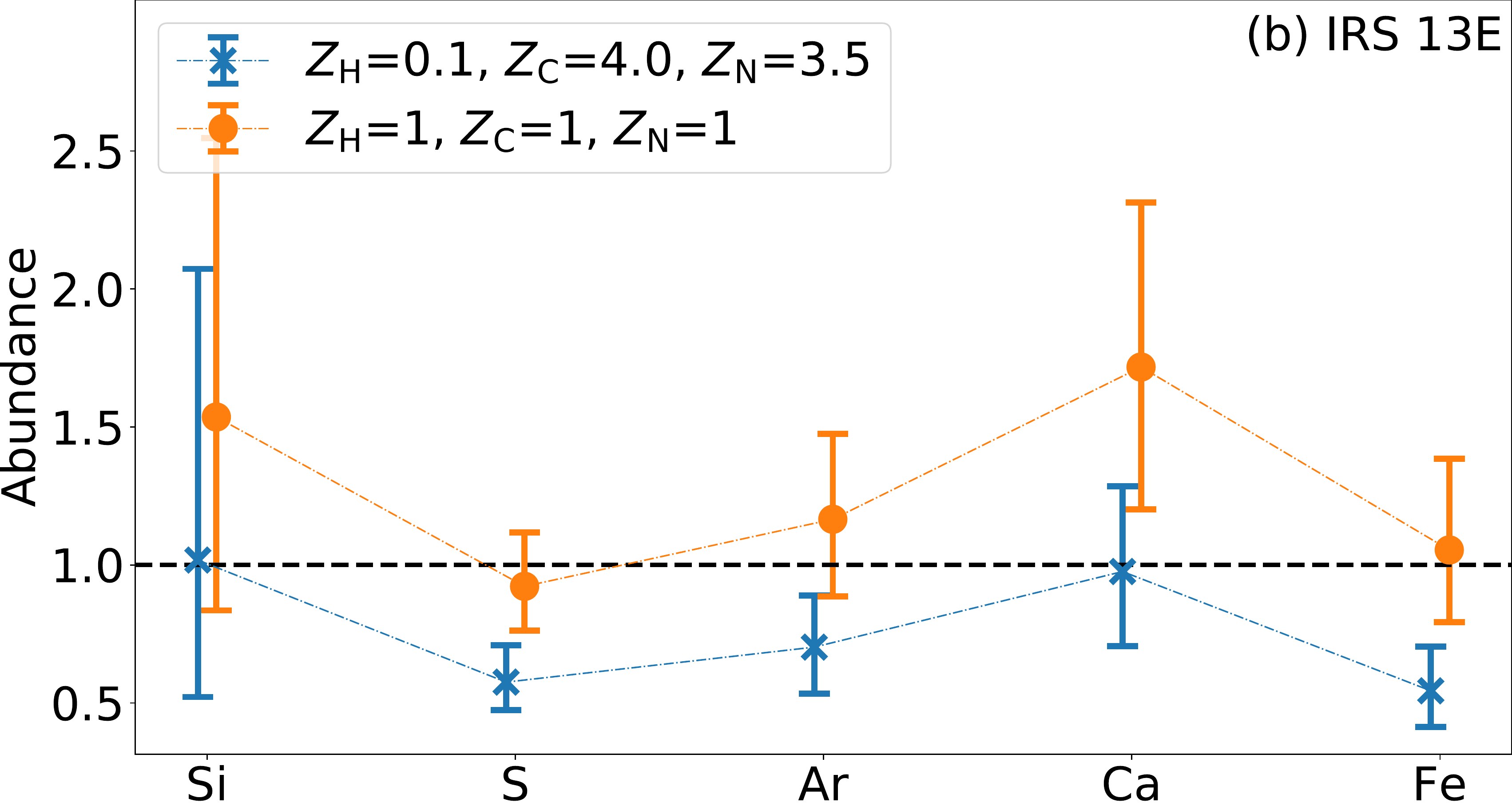}
	\includegraphics[width=0.48\textwidth]{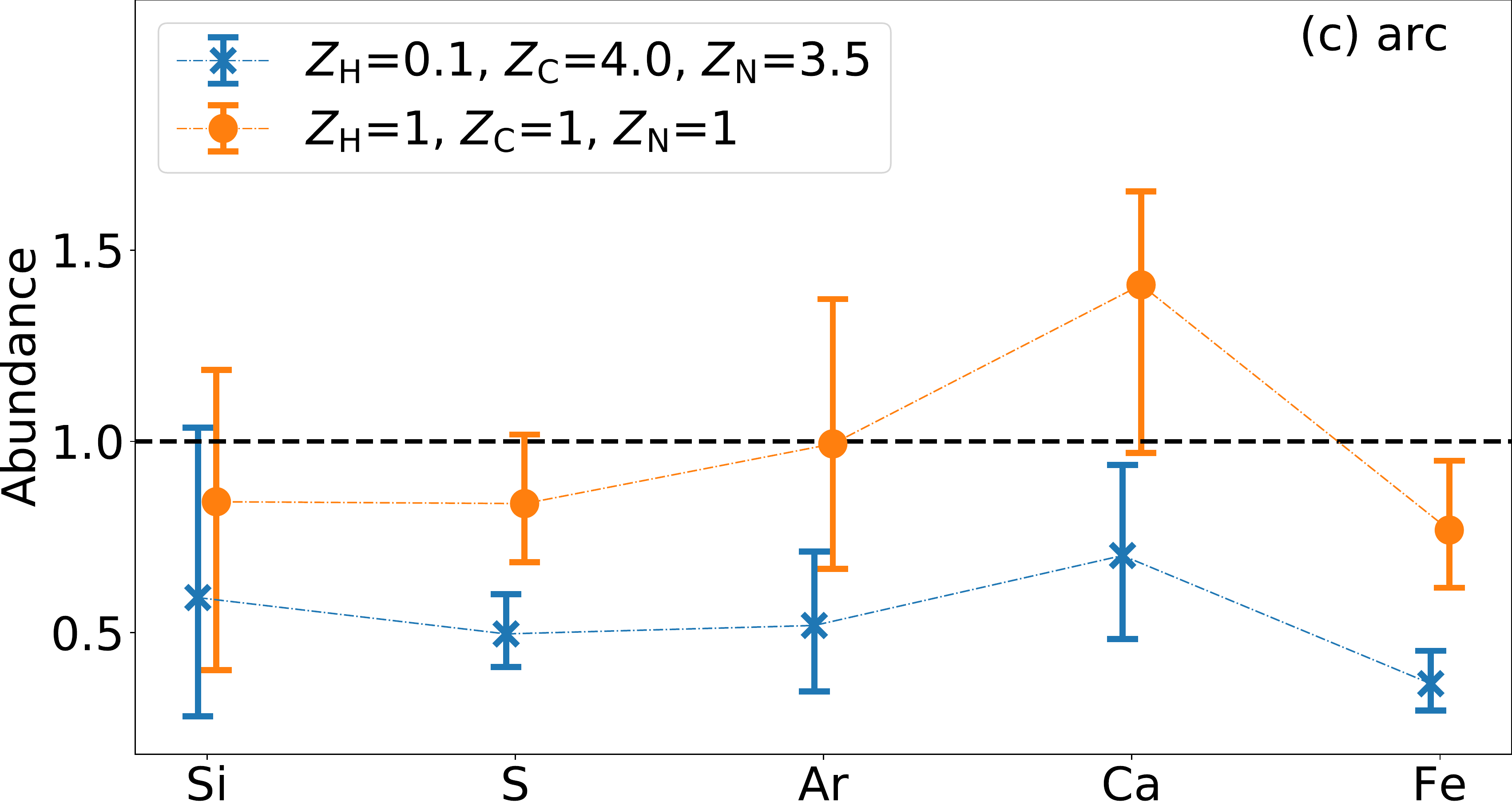}
	\includegraphics[width=0.48\textwidth]{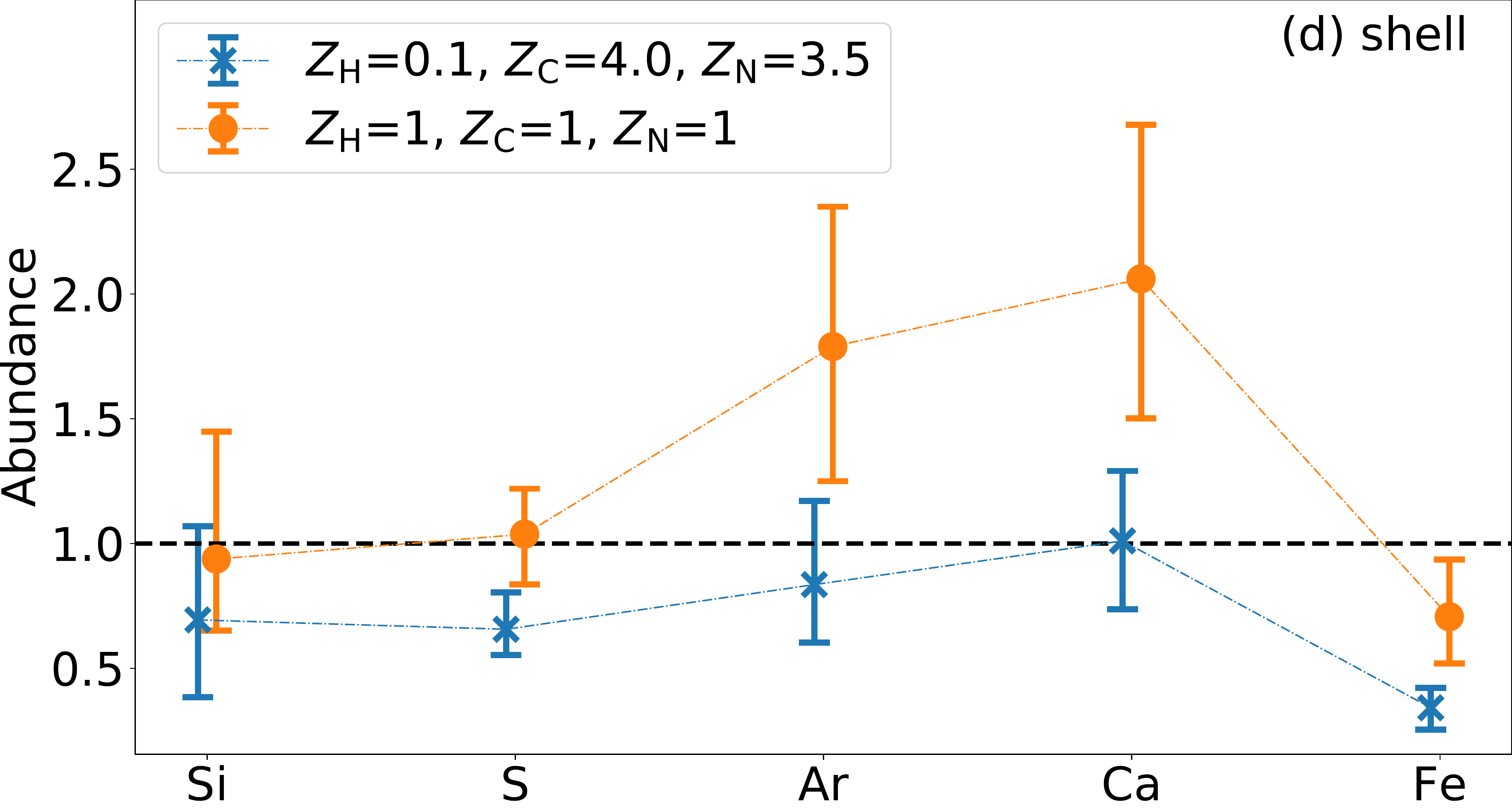}

	\caption{The best-fitted abundances of the heavy elements. The hydrogen-depleted and hydrogen-normal cases are shown in blue and orange, respectively, which are slightly shifted along the X-axis for clarity. The error bars are at the 90\% level. The horizontal black dashed line marks the solar value.}
	\label{fig:abundance_comparison1}
\end{figure*}

Fig.~\ref{fig:abundance_comparison1} shows the fitted heavy element abundances. 
We first examine
the physically more plausible case of $Z_{\rm H}$=0.1, $Z_{\rm C}$=4.0, $Z_{\rm N}$=3.5 (blue symbols).  
It can be seen that the four diffuse features share a similar abundance pattern, that is, all five heavy elements have a sub-solar or near-solar abundance, with the highest (near-solar) value found with Ca and lowest value ($\sim$0.3--0.5) found with Fe (or Si, in the case of the ridge).
S and Ar have intermediate values and show little variation among the four features. 
IRS 13E shows the highest Si abundance (near solar), but with a substantial uncertainty. This can be attributed  to its relatively large $N_{\rm H}$, which introduces severe absorption at low energies, especially the Si line at $\sim$1.8 keV.
The Fe abundance is $\sim$0.5 in the ridge and IRS 13E, and $\sim$0.35 in the arc and shell, with relatively small errors due to the strong lines of Helium-like Fe.

The case of $Z_{\rm H}$=$Z_{\rm C}$=$Z_{\rm N}$=1 (orange symbols) shows systematically higher abundances for all five elements, by a factor of $\sim 1.5-2$ compared to the hydrogen-depleted case. 
This can be understood as follows.
The total modelled spectrum mainly consists of H bremsstrahlung, He bremsstrahlung, C+N bremsstrahlung and the metal lines. 
The increase of the H bremsstrahlung, from 25.6\% to 255.9\% with respect to the He bremsstrahlung, more than compensates for the reduction of the C+N bremsstrahlung relative to the hydrogen-depleted case.
This thus requires a decrease in the hydrogen number density, and in turn leads to a higher metal abundance. 
Nevertheless, such a trend affects the different elements in a similar way, hence the overabundance pattern seen in a given feature or among the different features remains largely unchanged, as illustrated in Fig.~\ref{fig:abundance_comparison1}.

\begin{figure*}
	\centering
	\includegraphics[width=0.48\textwidth]{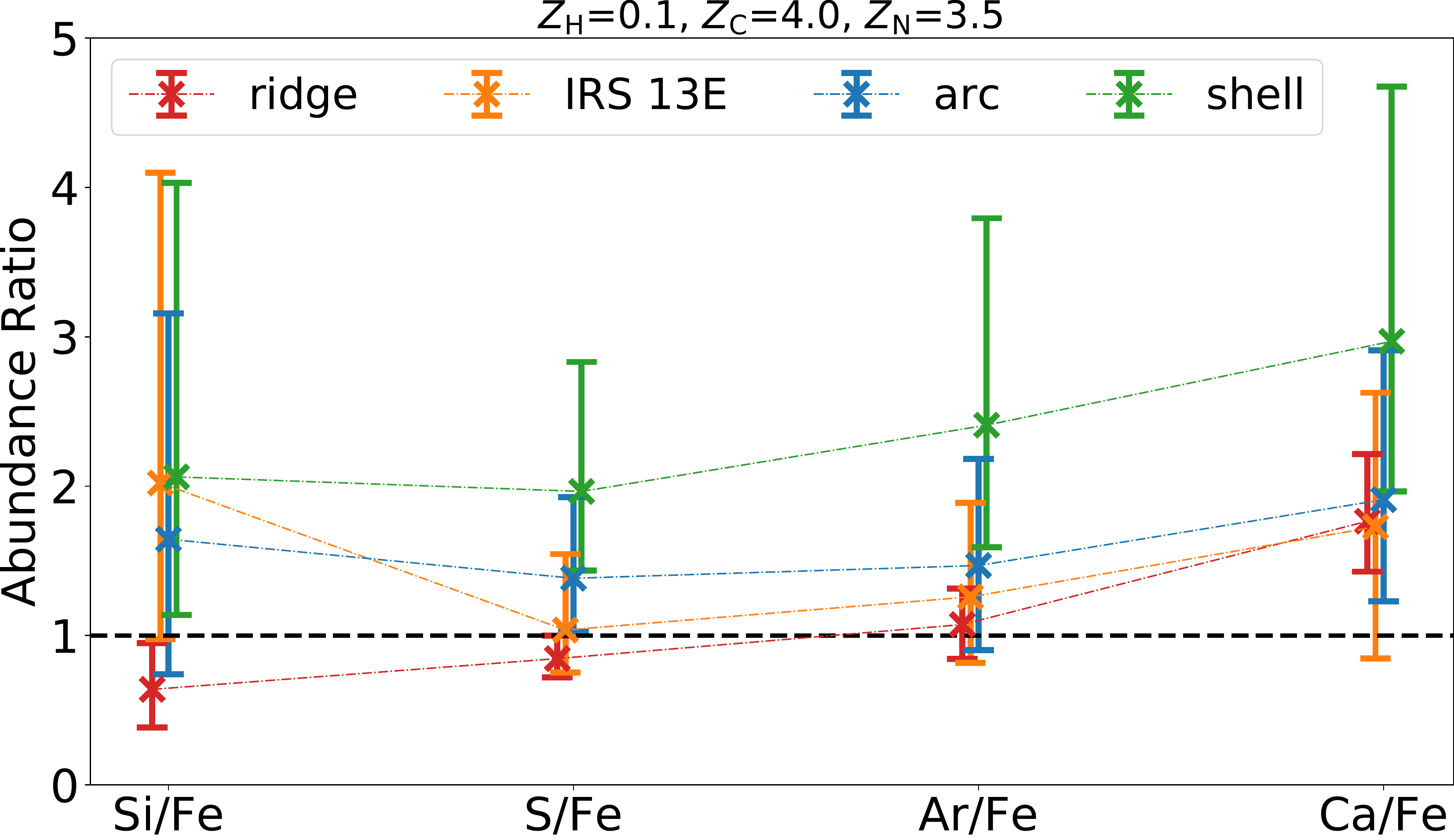}
	\includegraphics[width=0.48\textwidth]{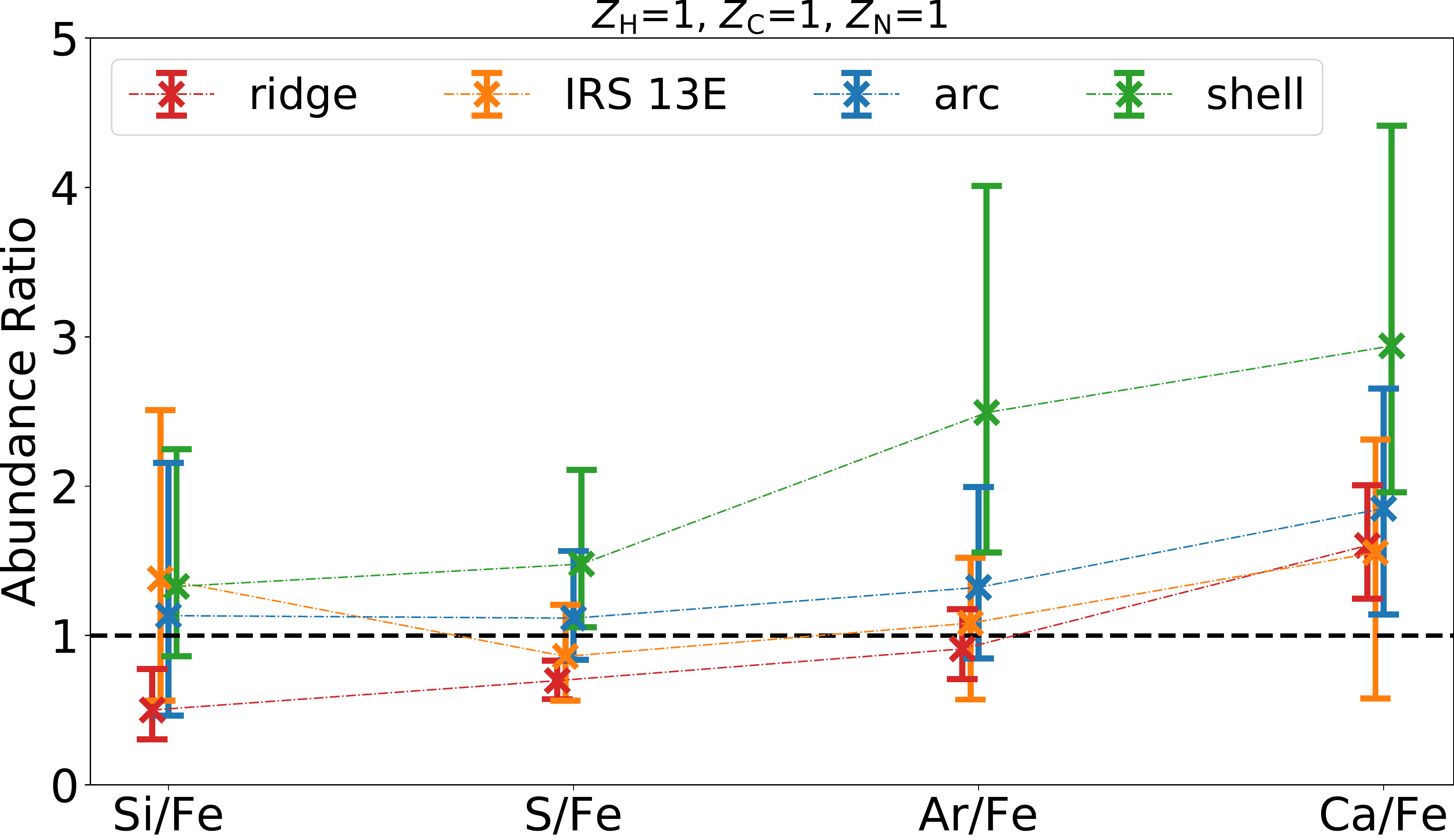}
	\caption{The $\alpha$/Fe abundance ratio. The left and right panels show the hydrogen-depleted and hydrogen-normal cases, respectively. The different color symbols represent different diffuse X-ray features, which are slightly shifted along the X-axis for clarity. The error bars are at the 90\% level. The horizontal black dashed line marks the solar value. }
	\label{fig:abundance_comparison2}
\end{figure*}

Fig.~\ref{fig:abundance_comparison2} further shows the $\alpha$ elements (Si, S, Ar and Ca) to Fe abundance ratios. The 90\% uncertainties in these ratios are evaluated by the Xspec function {\it fakeit} based on a bootstrapping method. 
Compared to the absolute abundance measurement, the $\alpha$/Fe ratio is less sensitive to the exact continuum level, as can be seen from the two panels showing the hydrogen-depleted and hydrogen-normal cases, respectively. 
Hence we focus on the hydrogen-depleted case, but noting that overall similar patterns are found in the hydrogen-normal case, except for an on-average somewhat lower ratio in Si/Fe and S/Fe.

The ridge shows a systematically lower $\alpha$/Fe ($\lesssim$1 solar) except for Ca/Fe, than the other three features. 
IRS 13E is relatively low in S/Fe and Ar/Fe, but is relatively high in Si/Fe. 
The latter could again be partly due to the higher absorption column density of IRS 13E.
The arc exhibits a relatively flat pattern, with values $\sim$1.5 in four ratios. 
The shell shows a systematically higher $\alpha$/Fe ($\sim$2), except for Si/Fe, when compared to the other three features.
This may imply a different physical origin, other than WR star winds, of the shell,
which will be further addressed in Section~\ref{subsec:shell}. 
We caution that in most cases the abundance ratios have a substantial measurement error.


\section{Discussion} \label{sec:discussion}
In Section~\ref{sec:model}, a phenomenological model is used to fit the diffuse X-ray spectra to constrain the abundance of heavy elements, including Si, S, Ar, Ca and Fe. Since the exact composition of light elements, including H, He, C and N, is uncertain and not directly constrained by the spectra, a degeneracy is introduced to the absolute abundance of the heavy elements.
If the underlying hot gas were hydrogen-depleted, which is consistent with the scenario that most, if not all, of the diffuse features originate from WR star winds,
the spectral fit finds a generally subsolar abundance for the heavy elements. If, instead, the light elements of the hot gas had a normal, solar-like abundance, the fitted abundances are systematically higher, ranging from solar to nearly twice solar. 
On the other hand, the $\alpha$/Fe abundance ratio, which is insensitive to the degeneracy brought by the light elements, is found to be supersolar ($\gtrsim 2$ solar in the shell, and about 1--1.5 solar in the other three features). 

\subsection{Potential bias in the spectral modeling}
\label{subsec:bias}
\begin{figure*}
	\centering
	\includegraphics[width=0.48\textwidth]{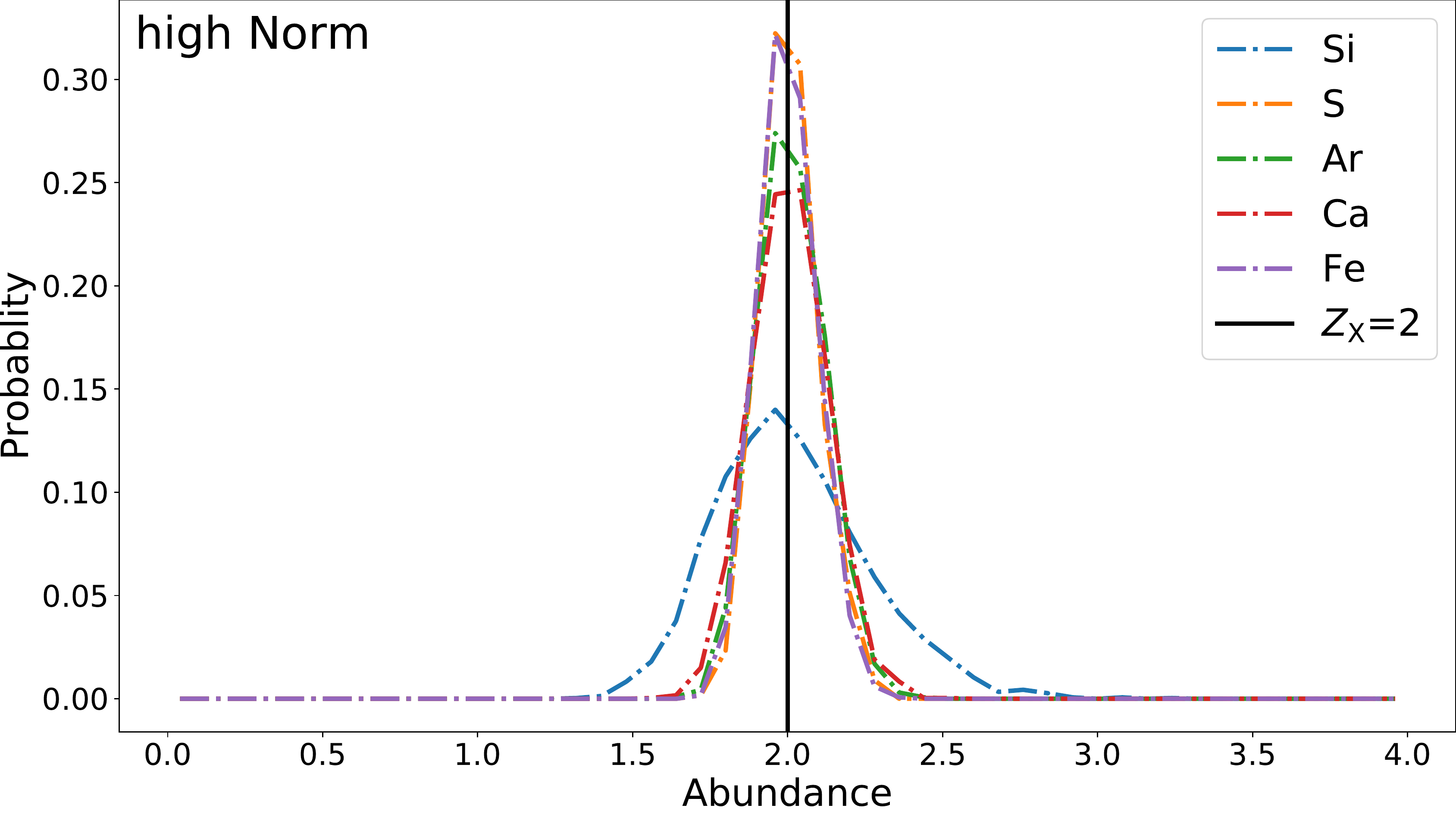}
    \hfill
	\includegraphics[width=0.48\textwidth]{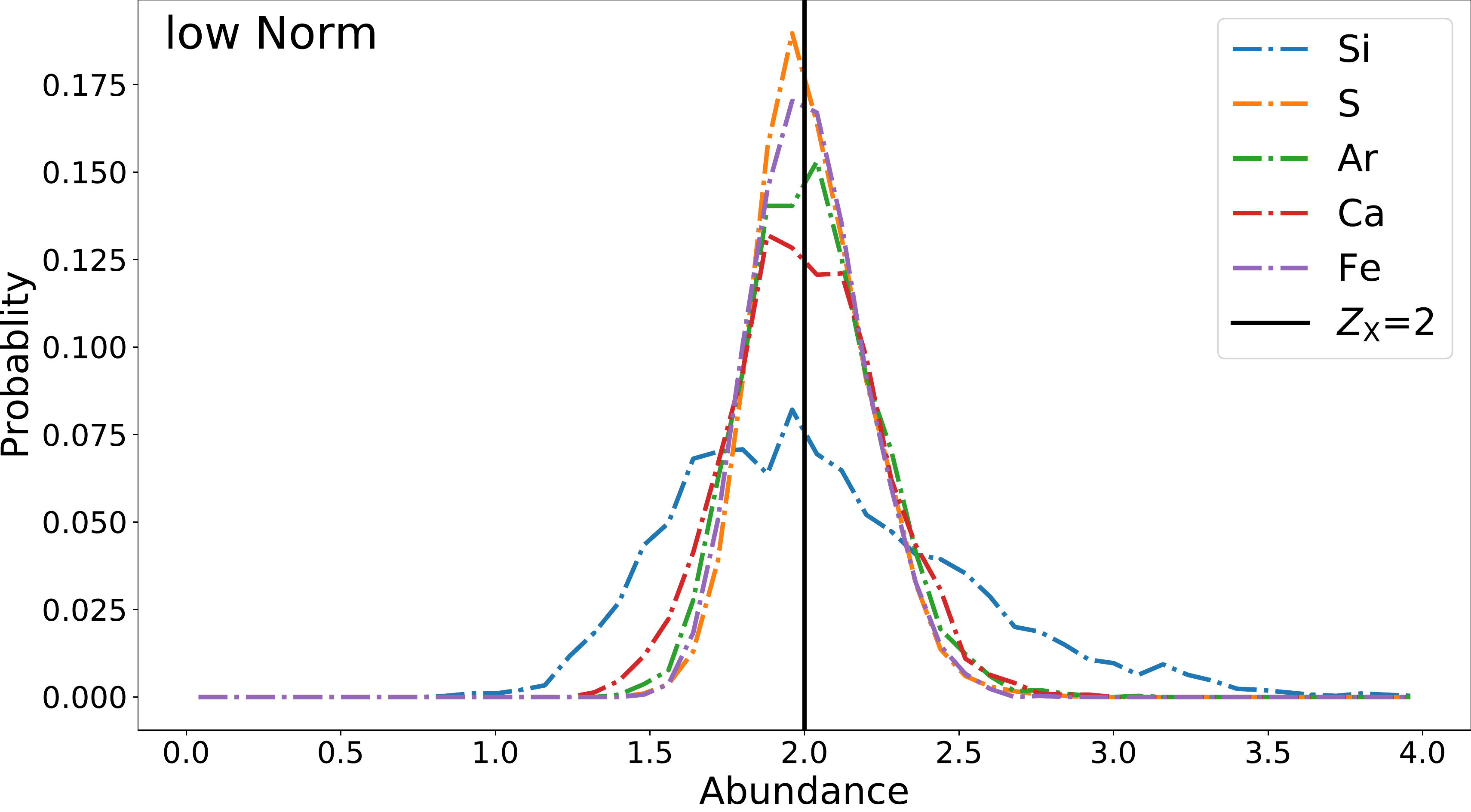}
 	\includegraphics[width=0.48\textwidth]{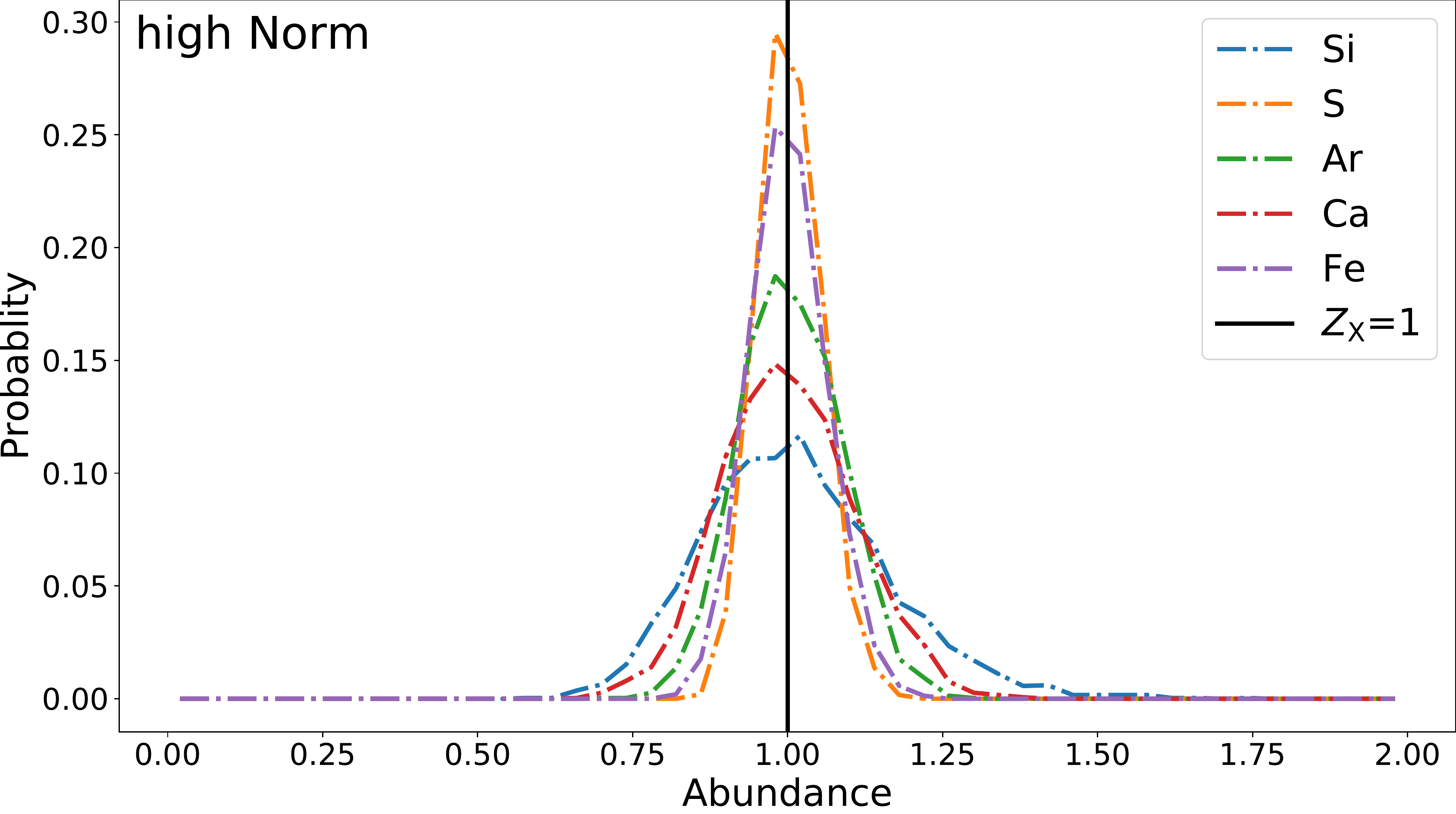}
	\hfill
	\includegraphics[width=0.48\textwidth]{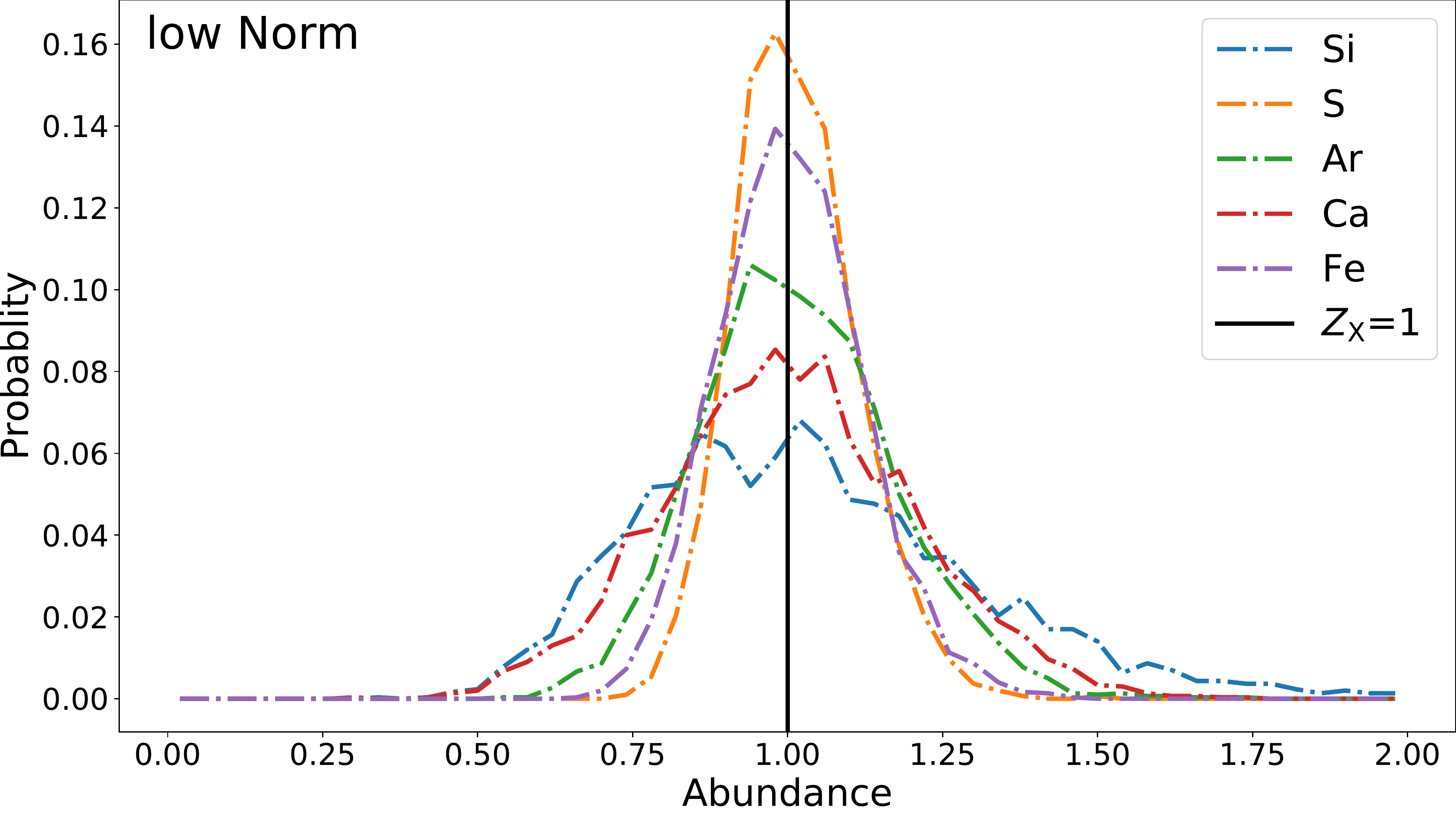}
	\includegraphics[width=0.48\textwidth]{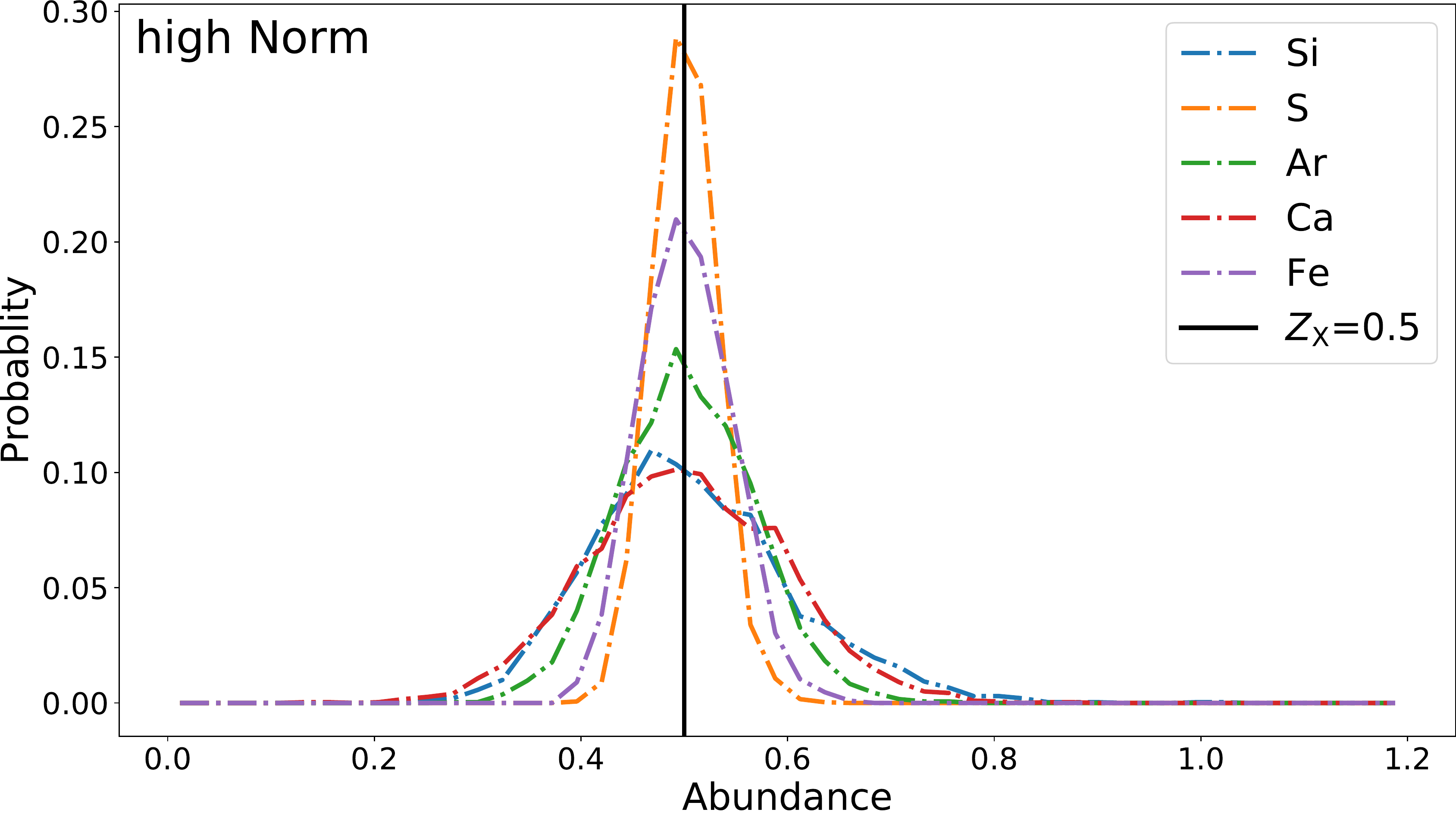}
	\hfill
	\includegraphics[width=0.48\textwidth]{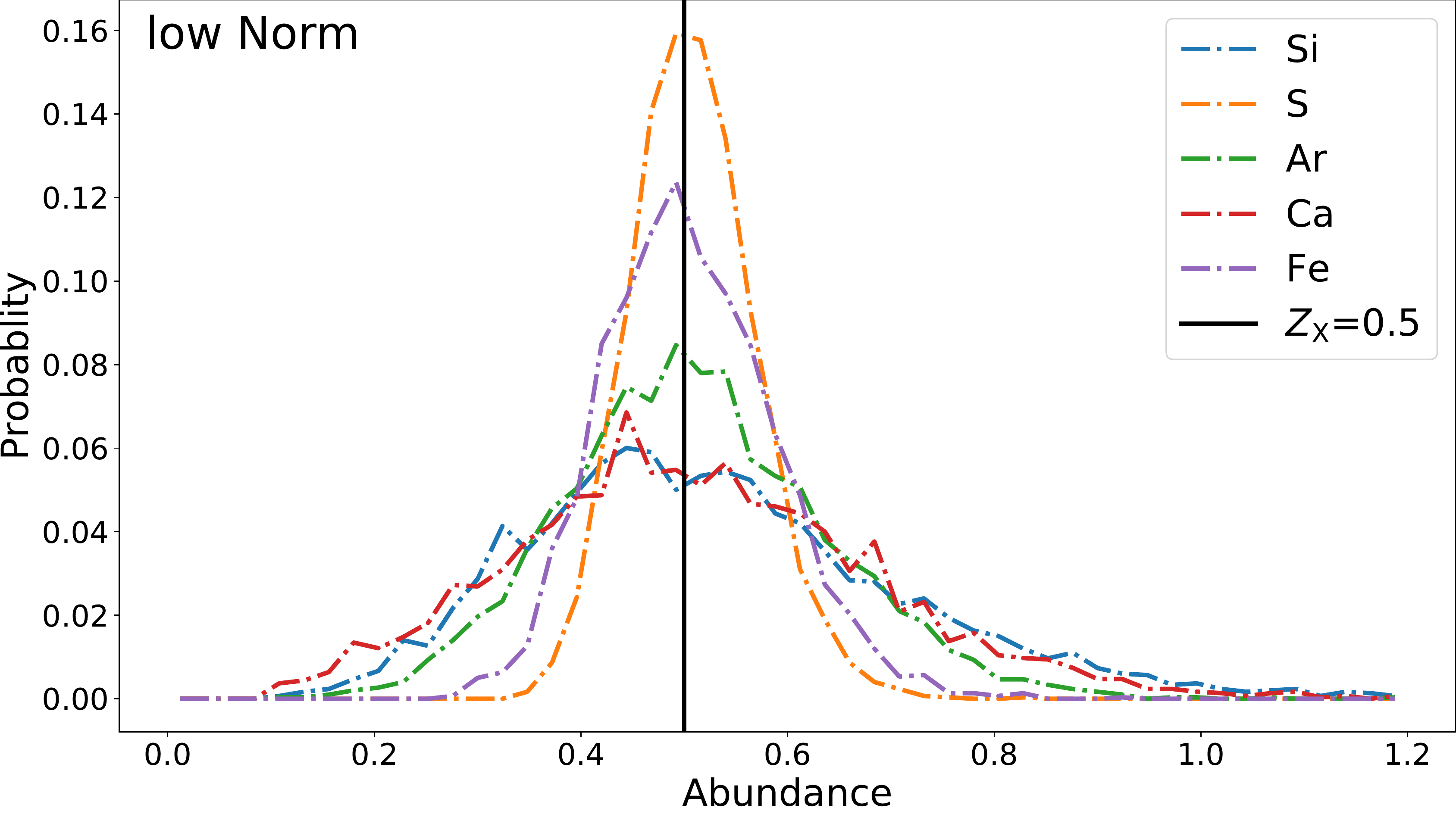}
	
	\caption{Probability distribution function of the fitted abundances from the simulated spectra test. The left and right panels are for cases of high and low normalization. The upper, middle and left panels are for an input abundance (same for all five elements) of 2, 1 and 0.5, respectively.}
	\label{fig:bias}
\end{figure*}

The best-fit parameters obtained from the spectral modeling, in particular the metal abundances, might be biased.
This is particularly relevant for moderate S/N spectra \citep[e.g.,][]{2009ApJ...693..822H}. 
To assess the potential bias, we perform tests with simulated spectra, which are generated from representative 2-T {\it vnei} models.
For the input parameters, we consider three different values of metal abundance (same for Si, S, Ar, Ca and Fe): 0.5, 1 and 2. 
The lighter elements have an abundance as in the hydrogen-depleted case.
The two temperatures are fixed at 0.8 keV and 4 keV, and the column density is set as $12\times10^{22} \rm~cm^{-2}$, which are 
representative of the actual best-fit values of the four features (Table \ref{tab:table1}). 
As for the normalization, we choose two sets of values to reflect the range of S/N in the real spectra: the high Norm case assumes $10^{-2}~\rm{cm^{-5}}$ for the low-temperature component and $10^{-3}~\rm{cm^{-5}}$ for the high-temperature component, and the low Norm case assumes $3\times10^{-3}~\rm{cm^{-5}}$ for the low-temperature component and $3\times10^{-4}~\rm{cm^{-5}}$ for the high-temperature component. 
For each combination of parameters, $3000\times3$ fake spectra are generated with the Xspec function {\it fakeit}.
Here the three-fold comes from the convolution of the ACIS-I, ACIS-G and ACIS-S spectral response files to mimic the joint-fit of the three groups in the actual cases. 
These fake spectra are then fitted with the same absorbed 2-T {\it vnei} model, allowing the column density, ionization timescale, metal abundances, temperatures and normalizations to vary as in the fit to the actual spectra. The resulting probability distribution functions (PDFs) of the fitted metal abundances are shown in Fig.~\ref{fig:bias}. 

For almost all parameter combinations, the PDFs follow a Gaussian-like shape, having a mean value consistent with the input abundance and a moderate scatter.
The largest scatter is seen in Si for any input abundance and norm, which is understood as Si having the lowest line energy ($\sim$ 1.8 keV) among all five elements, which suffers most from absorption.
For the other elements, the scatter is typically $\lesssim$0.1 in the high norm cases and $\lesssim$0.2 in the low norm cases, regardless of the input abundance. These are comparable to or smaller than the measurement errors.
We conclude from the simulated PDFs that the S/N of the observed spectra would not cause significant bias in the derived metal abundances. 

\subsection{Potential non-thermal components}
\label{subsec:nonthermal}
The spectral analysis in Section~\ref{sec:model} has assumed that the spectra of the diffuse features are purely dominated by thermal emission, in particular, the continuum is dominated by bremsstrahlung. 
However, since these diffuse features are understood as the manifestation of shock-heated gas, it is conceivable that the shocks have produced relativistic particles, which may also contribute to the observed X-ray spectra via non-thermal emissions such as synchrotron and inverse Compton scattering. 
Neglecting this potential non-thermal component might result in an overestimate of the bremsstrahlung component and in turn an underestimate of the metal abundance. 

However, it is not easy to determine the fractional contribution of the non-thermal component from a direct spectral decomposition, owing to the overall similar spectral shape in the keV range between the non-thermal and thermal continua.
Instead, we utilize multi-wavelength information to provide constraints on the non-thermal component, taking advantage of the fact that the same relativistic particles would produce measurable non-thermal emission outside the X-ray band. 
In particular, we make use of the Very Large Array (VLA) 5.5 GHz and 8.3 GHz observations \citep{2013ApJ...777..146Z}, as well as HESS, VERITAS and MAGIC gamma-ray (TeV) observations of the Galactic Center \citep{2016Natur.531..476H,2020A&A...642A.190M,2021ApJ...913..115A}. 
The VLA radio images have sufficiently high resolution to resolve the diffuse X-ray features.
However, in principle the emission from other physical components of the Galactic center, especially the sidelobes of Sgr A*, can contaminate the radio flux densities measured from the same regions used to extract the X-ray spectra. Hence the measured radio flux densities are effectively loose upper limits of the intrinsic radio emission from the diffuse features. 
The gamma-ray data, on the other hand, are of much lower angular resolutions ($\sim 0.1\deg$), hence the adopted gamma-ray fluxes of the Galactic center, while being concrete measurements, should also be treated as loose upper limits, and effectively the same limits, of the individual diffuse X-ray features. 

We note in passing that no IR counterpart is clearly present for any of the four features except for IRS 13E. In that case, the IR flux is dominated by thermal emission from dust, presumably produced in the colliding wind zone between the two WR stars \citep{2010ApJ...721..395F}. The K-band flux density of IRS 13E, $\sim85$ mJy \citep{2020ApJ...897..135Z}, is in fact much higher than the flux density of a non-thermal population estimated below.

\begin{figure*}
	\centering
	\includegraphics[width=0.48\textwidth]{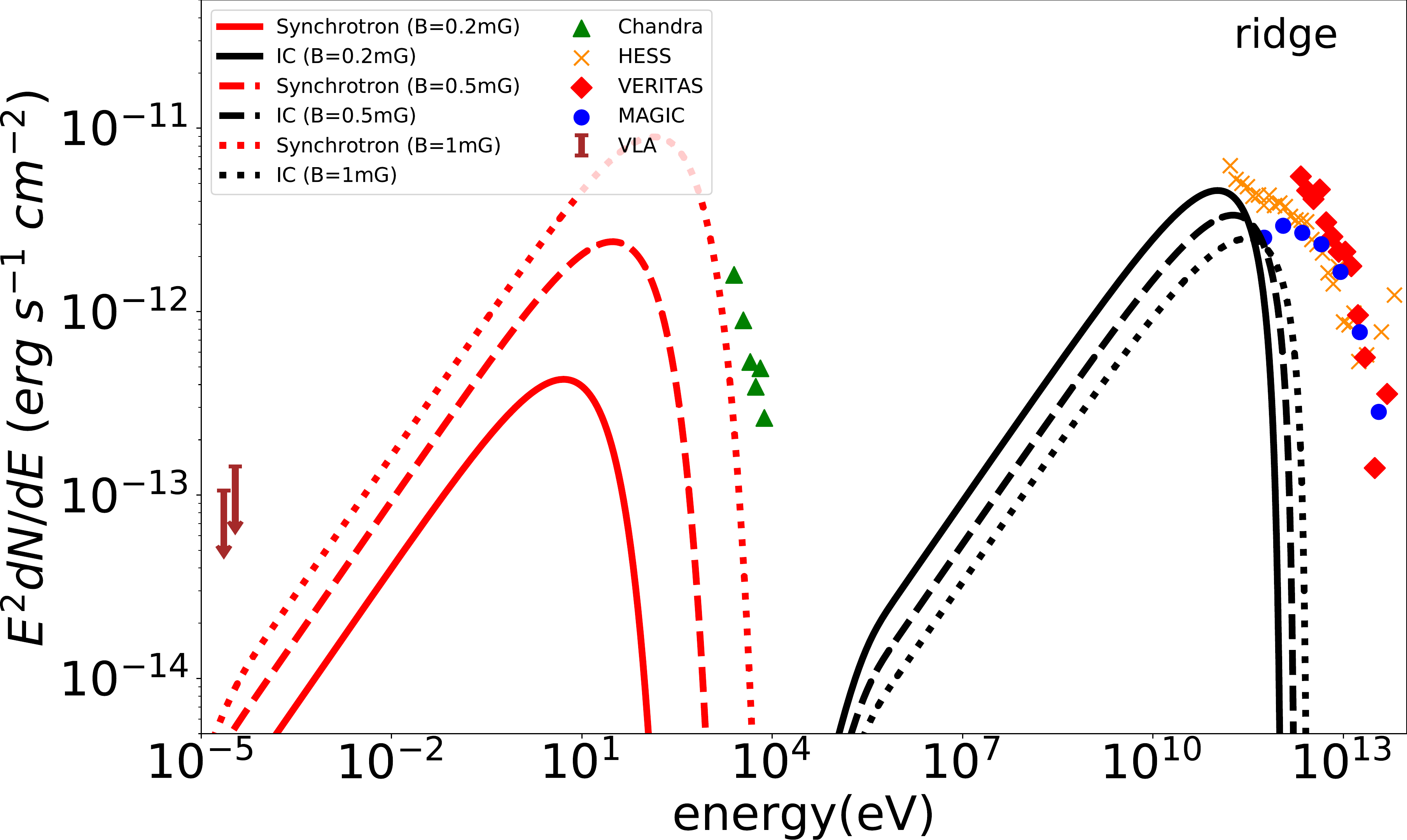}
	\includegraphics[width=0.48\textwidth]{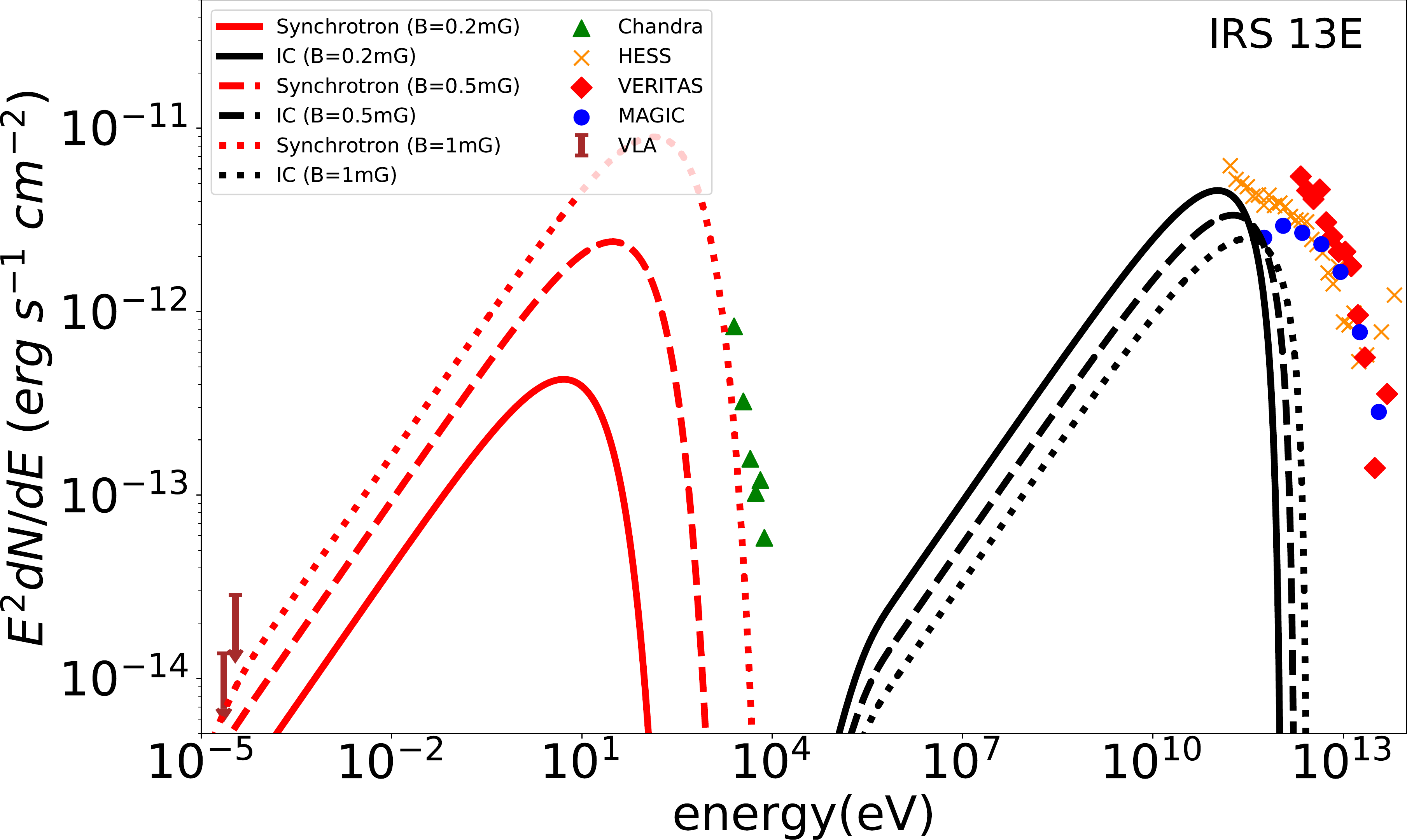}
	\includegraphics[width=0.48\textwidth]{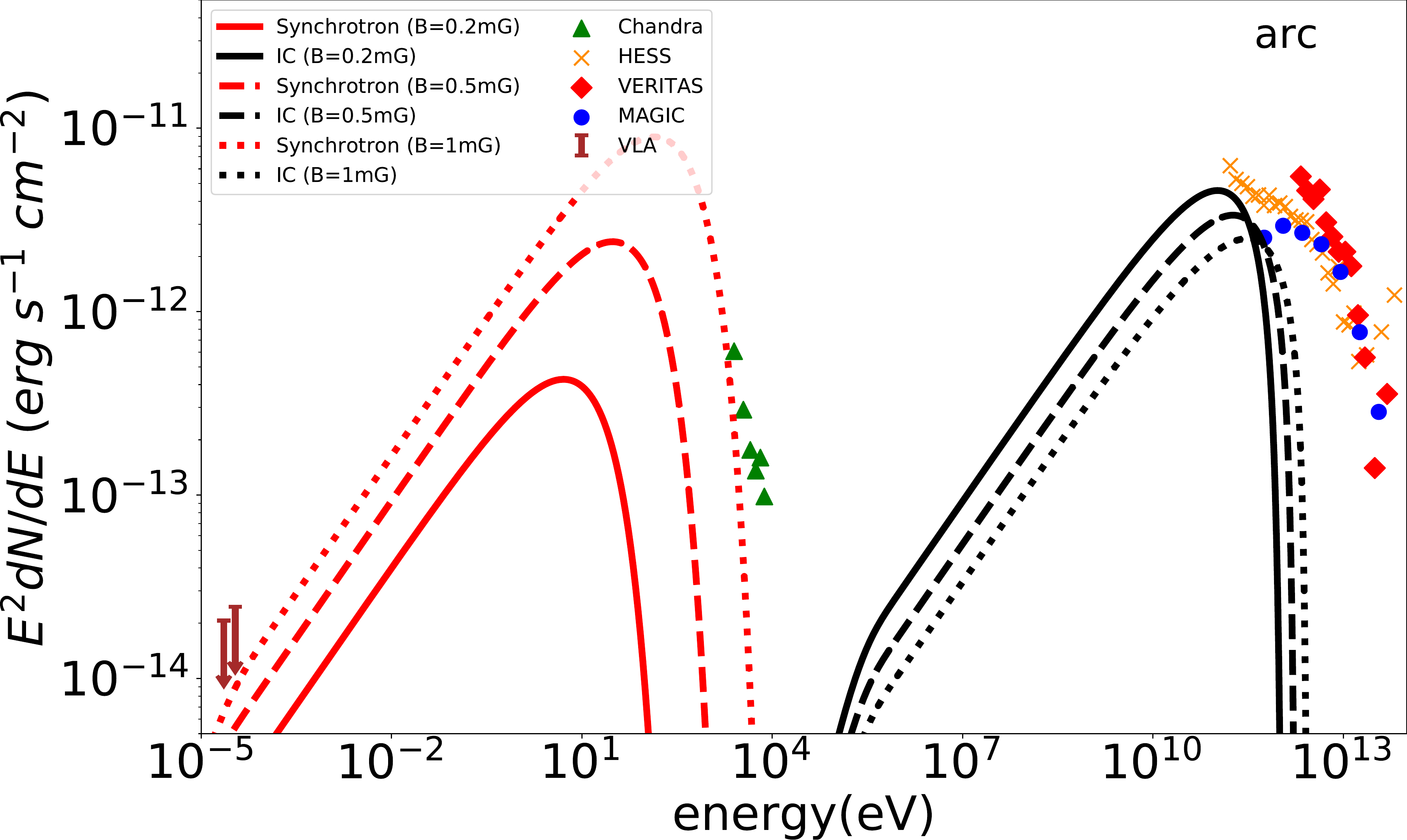}
	\includegraphics[width=0.48\textwidth]{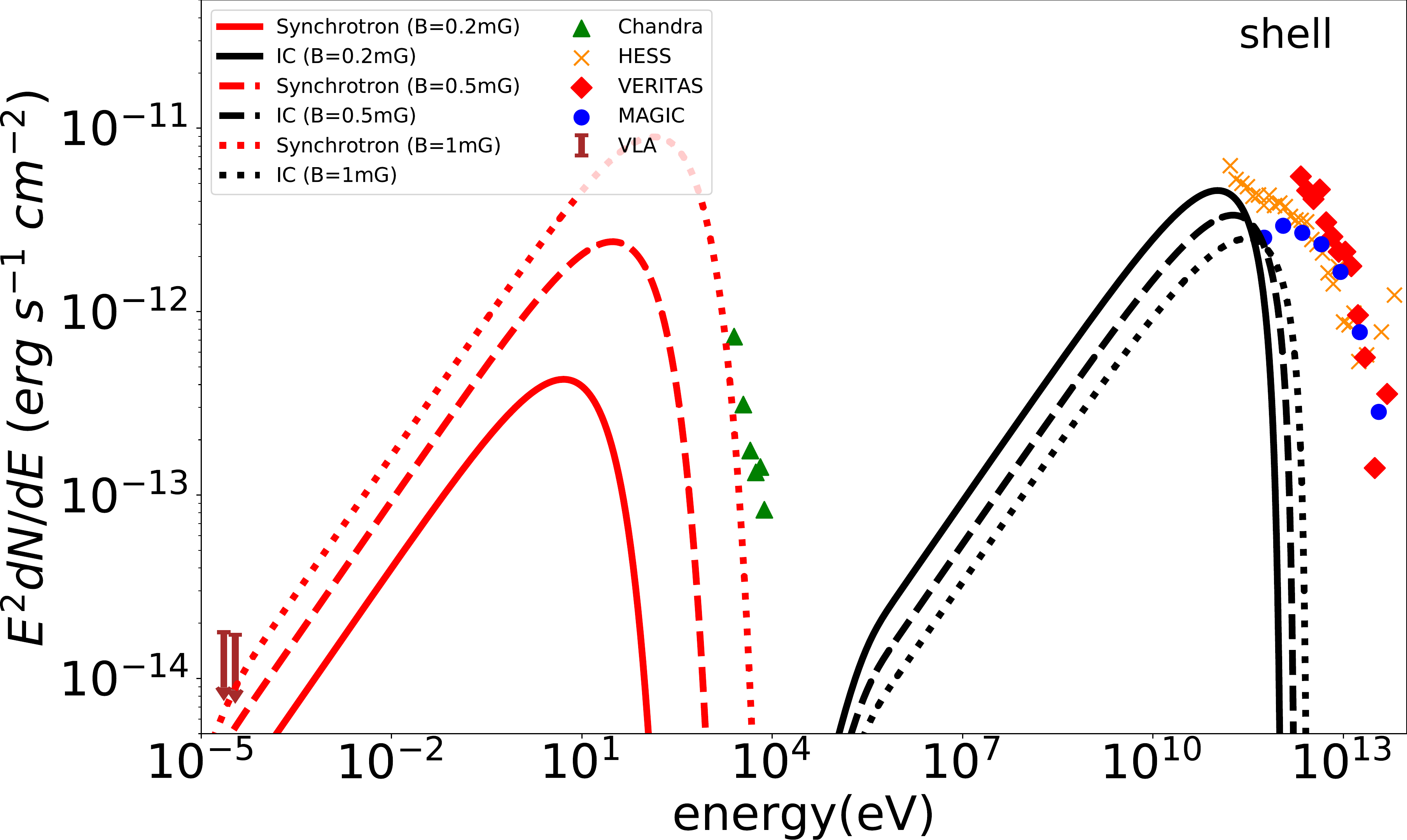}
	\caption{The broadband SED of the four diffuse X-ray features. The X-ray data (green triangles) have been corrected for foreground absorption. The radio upper limits (brown arrows) are derived from the VLA 5.5 and 8.3 GHz images of \citet{2013ApJ...777..146Z}. The TeV gamma-ray data are taken from HESS, VERITAS and MAGIC \citep{2016Natur.531..476H,2020A&A...642A.190M,2021ApJ...913..115A}, shown by the yellow crosses, red diamonds and blue dots, respectively.
	Three SEDs including synchrotron (red curve) and inverse Compton scattering (black curve) are modeled assuming different magnetic fields, which all predict an X-ray spectrum much lower than the one observed in all four features. See text for details.}
	\label{fig:SED}
\end{figure*}

The radio-X-ray-gamma-ray spectral energy distribution (SED) of the four features is shown in Fig.~\ref{fig:SED}, in which the X-ray data points have been corrected for the best-fit foreground absorption.
To contrast with this observed SED, we construct a two-component model, which consists of synchrotron radiation from a single population of relativistic electrons coupled with a uniform magnetic field, 
and inverse Compton (IC) scattering of the far-infrared Galactic center radiation field, which is dominated by dust-reprocessed radiation and represented by an energy density of $U_{\rm rad} = 3\times10^{-8}{\rm~erg~cm^{-3}}$ at a peak frequency of $6\times10^{12}$ Hz \citep{1992ApJ...387..189D}.
A synchrotron self-Compton (SSC) component has also been considered, but it turns out that this component is negligible compared to the external IC. 
To compute the synchrotron component, we adopt a power-law with an index of 2 for the electron energy distribution, and the maximum electron energy is 
determined by the requirement that the acceleration timescale of an electron at the shock front is shorter than the radiative cooling timescale. Following  \citet{2009MNRAS.393.1063T}, the acceleration timescale is
\begin{equation}\label{Eq1}
t_{\rm acc} = \frac{6\gamma m_e c^3}{e B v_s^2},
\end{equation}
where $v_s \approx 1000$ km~s$^{-1}$ is the shock velocity and $B$ is the magnetic field strength. The adopted value of the shock velocity is typical of the WR star winds. 
\citet{2010Natur.463...65C} estimated a typical value of $B \sim 100\ {\rm \mu G}$ based on energy equipartition between the magnetic field, X-ray emitting hot plasma, and turbulent gas across the Galactic center, whereas \citet{2020ApJ...890L..18T} suggested that a higher magnetic field $B \sim 200\ {\rm \mu G}$ can be found in the so-called non-thermal filaments. 
The field strength in the inner parsecs is uncertain but may be still higher.
Therefore, we examine three different magnetic field strengths: $B=0.2, 0.5, 1$ mG.
An electron with a Lorentz factor $\gamma$, moving in a radiation field with an energy density of $U_{\rm rad}$, will undergo synchrotron and IC radiation losses on a timescale,
\begin{equation}\label{Eq2}
t_{\rm loss} = \frac{3 m_e c}{4 \sigma_{\rm T} \gamma U_{\rm rad}},
\end{equation}
where $\sigma_{\rm T}$ is the Thompson scattering cross-section.
Thus $\gamma$ takes a maximum value
\begin{equation}\label{Eq2}
\gamma_{\rm max} = \sqrt{\frac{e B v_s^2}{4 \sigma_{\rm T} c^2 U_{\rm rad}}},
\end{equation}
which corresponds to a maximum electron energy of $1.3, 2, 3\ {\rm TeV}$, 
for $B = 0.2, 0.5, 1$ mG, respectively.

The model-predicted SEDs are calculated with a publicly available Python package {\it naima} \citep{naima} and shown in Fig.~\ref{fig:SED}. The normalization of the SEDs is maximally allowed such that they will not violate any observational upper limit in the radio and gamma-ray bands. It turns out that the tightest constraint comes from the MAGIC measurements, which, as emphasized before, are themselves loose upper limits. 
In all four cases, the predicted synchrotron flux has only a minor contribution to the measured X-ray flux and, due to the steep decay, affects mostly photon energies below 3 keV by up to $\sim10\%$ in the ridge and up to $\sim30\%$ in the arc, shell and IRS 13E.  
Hence we conclude that a potential non-thermal component in the diffuse X-ray features will not affect our results significantly, especially the abundance of Ar, Ca and Fe, which have emission lines at $> 3\ {\rm keV}$. 
Nevertheless, we caution that there is substantial uncertainty in our model parameters, and  a significant non-thermal contribution to the observed X-ray spectra cannot be definitely ruled out.

Conversely, Fig.~\ref{fig:SED} also suggests that the TeV gamma-ray emission from the innermost region of the Galactic center is unlikely to be dominated by the IC radiation from relativistic particles accelerated in the WR star winds.

\subsection{Comparison with IR metallicity measurements and implications for the Galactic center environment}\label{implication}

The inferred heavy element abundances of the diffuse X-ray features are to be contrasted with other metallicity measurements in the Galactic center. As mentioned in Section~\ref{sec:intro},
near-solar or moderately supersolar Fe abundance and $\alpha$/Fe have been found for a handful of red supergiants in the NSC \citep{2000ApJ...530..307C,2000ApJ...537..205R,2007ApJ...669.1011C,2009ApJ...694...46D}.
Recently, a hyper-velocity star (HVS) S5-HVS1 has been detected and investigated by the ${\rm S^{5}}$ collaboration \citep{2020MNRAS.491.2465K}.
This star is identified as an A-type star with a large radial velocity of $\sim1000\rm~km~s^{-1}$ and so far the only HVS confidently associated with the Galactic center. 
It is likely produced by tidal break-up of a tight binary when passing by Sgr A*.
\citet{2020MNRAS.491.2465K} inferred a $\sim$2 times solar abundance of Fe for S5-HVS1, based on its optical spectrum.
As for low-mass stars, most red giants share a solar or moderately supersolar metallicity and $\alpha$/Fe ratio, while a small fraction of red giants exhibits a subsolar metallicity \citep{2015ApJ...809..143D,2017MNRAS.464..194F}.
Typically, moderately supersolar metallicity has been documented with various types of stars in the NSC.
The same is true for the mini-spiral of ionized gas, with most results indicating a moderately supersolar abundance of Ne and Ar \citep{1980ApJ...241..132L,1994ApJ...430..236S,2002ApJ...566..880G}. 

At face value, the sub-solar or near-solar heavy element abundances of the diffuse X-ray features (Fig.~\ref{fig:abundance_comparison1}), assuming depletion of hydrogen inherited from the WR star winds, significantly deviates from the aforementioned results. 
The discrepancy with the supersolar metallicity found in the red supergiants, is particularly noteworthy, as these stars, with a typical age of $\sim 10-100$ Myr \citep{2017ars..book.....L}, may have been born in a star-forming episode relatively close to that of the WR stars (i.e., $\sim$4--6 Myr ago).
If the mini-spiral had originated from the {\it circumnuclear disk} (CND), as their spatial and kinematic properties suggest \citep{1996ARA&A..34..645M,2010RvMP...82.3121G}, the discrepancy in the metal abundances, at least of Ar, also seems to suggest that the star formation episode 4--6 Myr ago was not fed by the same molecular gas streamers that formed the CND  \citep{2017ApJ...847....3H}.

Assuming no or little depletion of hydrogen in the diffuse X-ray features can reduce the discrepancy with the IR measurements, but this raises a further question of how to reconcile with the standard picture of WR star wind-fed hot gas in the central parsec. We discuss two possible solutions in Section~\ref{subsec:Hsupply} and Section~\ref{subsec:dust} below.

On the other hand, the X-ray and IR observations agree on a supersolar
$\alpha$/Fe abundance ratio. 
This is consistent with a top-heavy initial mass function inferred for the YNC \citep{2010RvMP...82.3121G,2013ApJ...764..155L}.
In this case, more massive stars will lead to an enhanced birth rate of core-collapse supernovae (SNe), which in turn produce more $\alpha$-elements. 
However, the metal-enriched ejecta of any recent SN would have been pushed out from the central parsec by the strong WR star winds in a dynamical timescale of $\sim1000$ yr.
The fact that a highly supersolar metallicity is unseen in the hot gas indicates that no SN has occurred in the central parsec in the past millennium.  

\subsection{Supply of hydrogen by other stars}
\label{subsec:Hsupply}
We consider the possibility that hydrogen is supplied by mass loss from stars other than the WR stars.
Within the YNC, there exist $\sim$ 30 of O supergiants and O giants \citep{2006ApJ...643.1011P}, which produce stellar winds without significant hydrogen depletion. The mass-loss rate of the individual O (super)giants is about 1--2 orders of magnitude lower (i.e., $\sim 10^{-7}-10^{-6}\rm~M_\odot~yr^{-1}$; \citealp{2018A&A...620A..89N}) than that of the WR stars. Thus the collective mass loss from these O (super)giants would inject $\lesssim 10^{-5}\rm~M_\odot~yr^{-1}$ of hydrogen to the central parsec. After mixing with the hydrogen-depleted material from the WR stars ($\sim 3\times10^{-4}\rm~M_\odot~yr^{-1}$), the number fraction of hydrogen should increase by less than 10\%. 

Another potential supplier of hydrogen is the AGB stars.
At least 5 AGB stars are detected within the central parsec \citep{2020A&A...642A..81S}. 
The typical mass loss rate of individual AGB is $\lesssim 10^{-6}~\rm M_{\odot}~yr^{-1}$ \citep{2005ARA&A..43..435H}. Thus the collective injection of hydrogen from the AGB stars is at most comparable to that from the O (super)giants. 
Other stellar populations, such as red giants and main-sequence O/B stars, are also unlikely to be a substantial source of hydrogen, due to their much weaker stellar winds, even though they are more numerous in the central parsec. 
Therefore, we conclude that mixing of mass loss from other stellar populations is unlikely to significantly change the fractional abundance of hydrogen in the diffuse X-ray features. 



\subsection{Metal depletion into dust grains}
\label{subsec:dust}
Depletion of heavy elements into dust grains may be an alternative explanation for the relatively low metal abundances observed in the diffuse X-ray features.
The phenomenon of depletion into dust grains is commonly seen in the ISM. 
Several X-ray observations toward Galactic center sources, including SGR J1745-2900 and Swift J174540.7-290015, reveal depletion into dust grains in the ISM along the line-of-sight \citep{2016MNRAS.461.2688P,2017MNRAS.471.1819C}.
Moreover, at least in the case of IRS 13E, significant dust is present in the colliding wind zone between the two WR stars \citep{2010ApJ...721..395F,2020ApJ...897..135Z}, which is likely formed out of the post-shock gas \citep{1991MNRAS.252...49U}.
Recent JWST observations of the Wolf-Rayet binary WR140 also confirm the production of carbonaceous dust grains  \citep{2022NatAs.tmp..216L}.
The strength of depletion is related to measures of chemical affinity, such as condensation temperatures and atomic sticking probabilities. According to \citet{2009ApJ...700.1299J}, Fe undergoes the strongest depletion, followed by Si and S. 
Qualitatively, this might partly account for the observed subsolar abundances of these three elements and the supersolar $\alpha$/Fe ratios (except in the case of the ridge).

Once formed, the dust grains would be destroyed due to sputtering in the diffuse hot gas, within a timescale of $10^6 (a/{\rm {\mu}m)}({\rm cm^{-3}}/n_{\rm H})\rm~yr$ \citep{1996ApJ...457..244D}, where $a$ is the grain size and $n_{\rm H}$ is the hydrogen density. 
From the spectral analysis, we estimate a hydrogen density of $\lesssim 10^{2}\rm~cm^{-3}$ for the diffuse X-ray features. Thus the dust sputtering timescale is $\sim 10^{4}$ yr, which is longer than the time needed for the dust to escape from the central parsec, provided that the dust is entrained in the hot gas outflow with a radial velocity of $\sim 10^3\rm~km~s^{-1}$ \citep[e.g.,][]{2018MNRAS.478.3544R}.  
Thus the dust grains produced near the WR stars may not be completely destroyed after they diffuse into the hot gas.  

However, we do not expect significant depletion in Ar, which hardly reacts with other elements to form refractory compounds.
Since there is no evidence for an elevated abundance of Ar, compared to Si, S and Ca, the effect of depletion into dust grains is probably also small for these three elements. 
We note that \citet{1993ApJ...418..244L} also disfavored significant depletion of Fe into dust in the central parsec, unless the Fe abundance is 10 times or more above the solar value, which is highly unlikely.
Moreover, IRS 13E, which exhibits recently formed dust \citep{2010ApJ...721..395F}, shows no evidence of a lower abundance compared to the other three features.
Therefore, we conclude that depletion into dust grains is unlikely to significantly affect the observed abundance of the heavy elements in the diffuse X-ray features.

\subsection{Possible origin of the shell}
\label{subsec:shell}
The shell exhibits a significantly higher $\alpha$/Fe than the other three features, which deserves some remarks. 
Indeed, apparently no known massive stars (WR or O-type) are found near the shell (Fig.~\ref{fig:xray_image}), casting doubt on its relation to the stellar winds. 
Based on the morphology of the shell, we speculate that it is the expanding remnant of a stellar explosive event during which nucleosynthesis takes place. Heavy elements produced in such an event might be mixed with the ambient medium, leading to an elevated $\alpha$/Fe observed in the shell.
However, the case of a young supernova remnant can be ruled out. This is because, at a typical SN shock velocity of $\sim10^4\rm~km~s^{-1}$ -- the radius of the shell ($\sim$0.25 pc) implies a dynamical age of $\sim$25 yr -- both the morphology and X-ray flux of the shell should significantly vary on an annual timescale, which is not observed in the {\it Chandra} data. 
Therefore, we consider the possibility that the shell is the result of a weaker stellar explosion. 

One possibility is a nova remnant. The Galactic center is home to at least thousands of cataclysmic variables \citep{2018ApJS..235...26Z}. Hence nova explosions are expected to be frequent in this region, yet none has been conclusively observed. 
In the solar neighborhood, the best-known nova remnant, a shell-like feature, is associated with GK Per, a magnetic white dwarf binary, which underwent a classical nova outburst in 1901 \citep{1901MNRAS..61..337W}. At a distance of $\approx$ 470 pc \citep{1960stat.book..585M}, the X-ray-emitting nova shell of GK Per shows a circular symmetry with an angular size of $\approx$ 1 arcmin, corresponding to a physical size of 0.14 pc. \citet{2015ApJ...801...92T} used two {\it Chandra} observations taken in 2000 and 2013 to determine an expansion velocity of $300~{\rm km~s^{-1}}$. 
They also estimated the mass of the shocked plasma to be $2\times10^{-4}~{\rm M_{\odot}}$, 
which is more than 100 times lower than inferred for the Galactic center shell (Table~\ref{tab:table1}). 
Moreover, a shock velocity of $\gtrsim1000\rm~km~s^{-1}$ is required to produce the observed gas temperature of few keV in the shell.  
These points indicate that if the shell were indeed a nova remnant, the progenitor nova outburst must have had both an unusually high explosion energy and a large ejecta mass, which, however, is rather unlikely \citep{2005ApJ...623..398Y}.  

Another possible scenario is the remnant of a failed SN, which may be generated by certain massive progenitors \citep{2008ApJ...684.1336K} and leads to a type of intermediate luminosity red transients \citep{2020ApJ...897L..44T,2021A&A...654A.157C}.
Such stars undergo core collapse but produce no successful supernova. In this case, neutrino radiation leads to a decrease in the core mass. As a result, the outer part of the star will be overpressured and generate an outward sound pulse, which further forms a shock propagating throughout the star \citep{2018MNRAS.476.2366F}. A failed SN will have an explosion energy and ejecta mass intermediate between a nova and a supernova, thus might provide a promising explanation for the shell. 
We are undertaking a more quantitative study for this scenario.

\section{Summary} \label{sec:conclu}
In this work, we have utilized ultra-deep {\it Chandra} X-ray observations to investigate the heavy element abundances of four prominent diffuse X-ray features located in the central parsec of the Galaxy. These features are thought to be the manifestation of shock-heated hot gas, which is supported by their 1.5--9 keV spectra exhibiting strong emission lines from Si, S, Ar, Ca and Fe. A two-temperature NEI model is employed to derive the element abundances and other physical quantities of the underlying hot gas. 

A degeneracy is inevitably introduced to the absolute abundance of the heavy elements, due to uncertainties in the composition of light elements, in particular, H, C and N.
If the underlying hot gas were hydrogen-depleted, as would be expected for a standard scenario in which the hot gas is dominated by WR star winds, the spectral fit finds a generally subsolar abundance for the heavy elements.
If, instead, the light elements have a solar-like abundance, the fitted abundances of the heavy elements range from solar to nearly twice solar. The $\alpha$/Fe abundance ratio is found to be supersolar and is insensitive to the exact composition of the light elements.
These results are robust against potential biases due to either a moderate spectral S/N or the presence of non-thermal components.

The sub-solar heavy-element abundances of the hot gas, under the assumption of hydrogen depletion in the WR star winds, are systematically lower than previous IR measurements of stellar metallicity and warm gas metallicity. We find that this discrepancy cannot be resolved by invoking an extra supply of hydrogen by other stars. Nor can it be fully understood in terms of metal depletion into dust. 

This study has provided one of the first observational constraints on the heavy element abundances of the hot gas, and indirectly for the massive stars, in the NSC.
Extending the measurement of heavy element abundances to the IR band and to other regions (e.g., the Arches and Quintuplet star clusters) in the Galactic center, will be important to further our understanding of the origin and diversity of the most massive and youngest stars in this unique environment of the Galaxy.

\section*{Acknowledgements}
This work is supported by the
National Natural Science Foundation of China (grants 11873028 and 12225302).
M.M. gratefully acknowledges support from NASA grant GO1-22138X to UCLA.
The authors wish to thank Zhao Su for helpful discussions on nova remnant and failed SN. 

\section*{Data Availability}
The data underlying this article will be shared on reasonable request to the corresponding author.



\bibliographystyle{mnras}
\bibliography{main} 





\bsp	
\label{lastpage}
\end{document}